\documentclass[
  aps,
  prl,
  reprint,
  superscriptaddress,
  amsmath,
  amssymb,
  longbibliography,
  floatfix
]{revtex4-2}
\usepackage[T1]{fontenc}
\usepackage{amsmath}
\usepackage{graphicx}
\usepackage{xcolor}
\usepackage{bm}
\usepackage[colorlinks,citecolor=blue]{hyperref}
\begin{document}

\title{Programmable cavity-enhanced telecom quantum memory in thin-film lithium niobate}

\author{Chengdong Yang}
\affiliation{National Laboratory of Solid-state Microstructures, School of Physics, College of Engineering and Applied Sciences, Collaborative Innovation Center of Advanced Microstructures, Jiangsu Physical Science Research Center, Nanjing University, Nanjing 210093, China}
\author{Hanwen Guo}
\affiliation{National Laboratory of Solid-state Microstructures, School of Physics, College of Engineering and Applied Sciences, Collaborative Innovation Center of Advanced Microstructures, Jiangsu Physical Science Research Center, Nanjing University, Nanjing 210093, China}
\author{Yu-Yang An}
\affiliation{National Laboratory of Solid-state Microstructures, School of Physics, College of Engineering and Applied Sciences, Collaborative Innovation Center of Advanced Microstructures, Jiangsu Physical Science Research Center, Nanjing University, Nanjing 210093, China}
\author{Qian He}
\affiliation{National Laboratory of Solid-state Microstructures, School of Physics, College of Engineering and Applied Sciences, Collaborative Innovation Center of Advanced Microstructures, Jiangsu Physical Science Research Center, Nanjing University, Nanjing 210093, China}
\author{Chi Lu}
\affiliation{National Laboratory of Solid-state Microstructures, School of Physics, College of Engineering and Applied Sciences, Collaborative Innovation Center of Advanced Microstructures, Jiangsu Physical Science Research Center, Nanjing University, Nanjing 210093, China}
\author{Ziheng Jiang}
\affiliation{National Laboratory of Solid-state Microstructures, School of Physics, College of Engineering and Applied Sciences, Collaborative Innovation Center of Advanced Microstructures, Jiangsu Physical Science Research Center, Nanjing University, Nanjing 210093, China}
\author{Yan-Qing Lu}
\affiliation{National Laboratory of Solid-state Microstructures, School of Physics, College of Engineering and Applied Sciences, Collaborative Innovation Center of Advanced Microstructures, Jiangsu Physical Science Research Center, Nanjing University, Nanjing 210093, China}
\author{Shining Zhu}
\affiliation{National Laboratory of Solid-state Microstructures, School of Physics, College of Engineering and Applied Sciences, Collaborative Innovation Center of Advanced Microstructures, Jiangsu Physical Science Research Center, Nanjing University, Nanjing 210093, China}
\author{Xiao-Song Ma}
\thanks{Corresponding author: Xiaosong.Ma@nju.edu.cn}
\affiliation{National Laboratory of Solid-state Microstructures, School of Physics, College of Engineering and Applied Sciences, Collaborative Innovation Center of Advanced Microstructures, Jiangsu Physical Science Research Center, Nanjing University, Nanjing 210093, China}
\affiliation{Synergetic Innovation Center of Quantum Information and Quantum Physics, University of Science and Technology of China, Hefei, Anhui 230026, China}
\affiliation{Hefei National Laboratory, Hefei 230088, China}
\date{\today}

\begin{abstract}
Spectrally multiplexed telecom quantum networks require quantum memories combining efficient storage with programmable frequency addressing. An integrated implementation should therefore unite a native telecom transition, efficient storage, and fast on-chip spectral control. Here we demonstrate a cavity-enhanced memory in an isotopically purified $^{167}\mathrm{Er}^{3+}$-doped thin-film lithium niobate microring. Long-lived hyperfine shelving states enable persistent, high-contrast atomic frequency comb preparation with a single-component lifetime of $277.6(52.6)$~s, while cavity impedance matching yields $23.3(5)\%$ on-chip efficiency for 100-ns storage. The intrinsic electro-optic response enables frequency-selective storage and routing at rates up to 20~MHz. We further store and retrieve time-energy-entangled telecom photons, violating an entanglement-witness bound by more than 11 standard deviations. Our results establish erbium-doped thin-film lithium niobate as a programmable light--matter interface for spectrally multiplexed quantum networks.
\end{abstract}

\maketitle

Quantum memories are key components of quantum networks, where they synchronize probabilistic photonic processes and enable long-distance entanglement distribution~\cite{briegelQuantumRepeatersRole1998,duanLongdistanceQuantumCommunication2001,kimbleQuantumInternet2008,sangouardQuantumRepeatersBased2011,wehnerQuantumInternetVision2018}. In spectrally multiplexed quantum networks~\cite{sinclairSpectralMultiplexingScalable2014,saglamyurekMultiplexedLightmatterInterface2016,seriQuantumStorageFrequencyMultiplexed2019}, a practical memory should not only store photons with long storage times and high efficiency, but also be spectrally tunable to interface with fixed-frequency sources, support channel-selective operation, and allow programmable spectral access. Rare-earth-ion-doped solids~\cite{lvovskyOpticalQuantumMemory2009,leiQuantumOpticalMemory2023} are promising candidates because they offer narrow optical transitions embedded in a broad inhomogeneous linewidth~\cite{thielRareearthdopedMaterialsApplications2011}, long-lived spin states~\cite{zhongOpticallyAddressableNuclear2015,maOnehourCoherentOptical2021} and mature echo-based storage protocols~\cite{afzeliusMultimodeQuantumMemory2009}.

Erbium is particularly attractive for fiber-based quantum networks because its optical transition lies directly in the low-loss telecom band near 1.5~$\mu$m~\cite{reisererColloquiumCavityenhancedQuantum2022}. Early demonstrations with naturally abundant erbium ensembles established storage of weak coherent states~\cite{lauritzenTelecommunicationWavelengthSolidStateMemory2010} and non-classical light~\cite{saglamyurekQuantumStorageEntangled2015,saglamyurekMultiplexedLightmatterInterface2016,askaraniStorageReemissionHeralded2019,zhangTelecombandintegratedMultimodePhotonic2023}, while more recent work with isotopically purified $^{167}\text{Er}^{3+}$ ensembles has exploited long-lived hyperfine shelving states~\cite{rancicCoherenceTimeSecond2018} to enable more coherent~\cite{craiciuNanophotonicQuantumStorage2019,rakonjacLongSpinCoherence2020,stuartInitializationProtocolEfficient2021,jiangQuantumStorageEntangled2023,anQuantumTeleportationTelecom2025} and functional memories~\cite{craiciuMultifunctionalOnchipStorage2021,liuOnDemandStoragePhotonic2022,liEfficientStorageMultidimensional2025}. Single erbium ions have also been employed as telecom single-photon sources~\cite{dibosAtomicSourceSingle2018,xiaTunableMicrocavitiesCoupled2022,ulanowskiSpectralMultiplexingTelecom2022,yangControllingSingleRare2023,ourariIndistinguishableTelecomBand2023,yuFrequencyTunableCavityEnhanced2023,huangStarkTuningTelecom2023} and for establishing spin--photon interfaces~\cite{gritschOpticalSingleshotReadout2025,uysalSpinPhotonEntanglementSingle2025a,ulanowskiCavityenhancedOpticalReadout2026}. Yet several requirements remain unmet within one device: high efficiency demands persistent high-contrast spectral preparation~\cite{askaraniPersistentAtomicFrequency2020,baryaUltraHighQTunable2025a} and cavity enhancement to compensate for the limited single-pass optical depth~\cite{afzeliusImpedancematchedCavityQuantum2010,moiseevEfficientMultimodeQuantum2010}; networking demands electrical frequency control to lock onto fixed-frequency sources and route heralded photons~\cite{sinclairSpectralMultiplexingScalable2014,assumpcaoThinFilmLithium2024}; and scalability demands an integrated platform~\cite{pelucchiPotentialGlobalOutlook2021}.

Lithium niobate (LN) is an attractive host for meeting these demands because it combines mature integrated photonics with a strong Pockels effect~\cite{wangIntegratedLithiumNiobate2018,boesLithiumNiobatePhotonics2023}. Although long recognized as a versatile host for rare-earth-ion quantum memories~\cite{saglamyurekBroadbandWaveguideQuantum2011,jiangRareEarthimplantedLithium2019,duttaAtomicFrequencyComb2023a}, earlier erbium-doped lithium-niobate devices relied on Ti-diffused or laser-written waveguides with weak optical confinement or material-induced decoherence~\cite{sinclairPropertiesRareEarthIonDopedWaveguide2017,askaraniStorageReemissionHeralded2019,zhangTelecombandintegratedMultimodePhotonic2023}, limiting storage efficiency below 3\%. Thin-film lithium niobate (TFLN) doped with isotopically purified $^{167}\text{Er}^{3+}$ provides a route beyond these limitations: high-$Q$ nanophotonic resonators can be impedance-matched to the erbium ensemble to enhance the effective light--matter interaction~\cite{afzeliusImpedancematchedCavityQuantum2010,moiseevEfficientMultimodeQuantum2010,duttaIntegratedPhotonicPlatform2020}, while the long-lived hyperfine shelving states of $^{167}\text{Er}^{3+}$ enable persistent, high-contrast spectral preparation. These advantages convert into high storage efficiency---all in a single chip that eliminates the need for external cavity stabilization. Integrated electrodes on the same waveguide further provide direct electrical control of the cavity resonance, which is especially valuable at millikelvin temperatures where slow tuning methods such as gas condensation are hard to scale~\cite{craiciuNanophotonicQuantumStorage2019}. An integrated erbium platform that simultaneously provides efficient cavity-enhanced storage and fast electrical programmability has therefore not yet been realized.

Here we realize such a platform in an isotopically purified $^{167}\text{Er}^{3+}$:TFLN microring, demonstrating cavity-enhanced AFC storage at $23.3(5)\%$ on-chip efficiency for 100-ns delay, electro-optic (EO) frequency-selective routing with $<10^{-1}$ crosstalk, and storage of time-energy-entangled telecom photons. These results establish erbium-doped thin-film lithium niobate as a programmable telecom-band light--matter interface for spectrally multiplexed quantum networks.

\section*{Results}

\subsection*{Device Design and Optical Characterization}

\begin{figure*}[t]
\centering
\includegraphics[width=\textwidth]{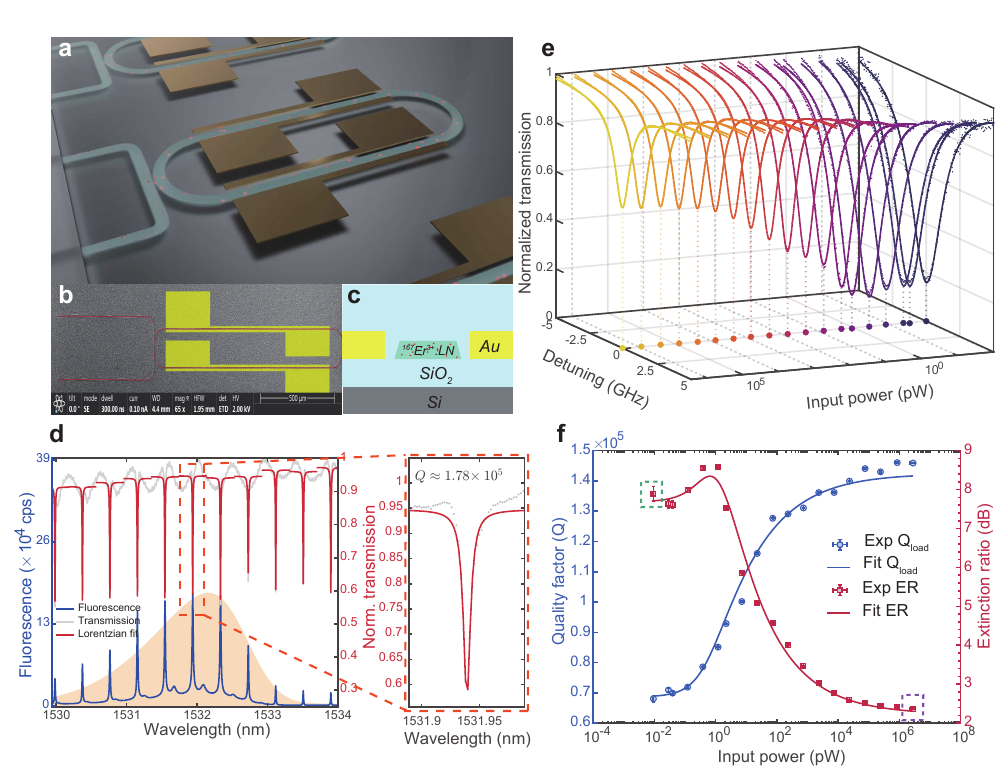}
\caption{\textbf{Device layout and optical characterization of the thin-film lithium niobate (TFLN) quantum memory.} \textbf{a}, Schematic of the $^{167}\text{Er}^{3+}$-doped TFLN device. \textbf{b}, Scanning electron microscopy image of the device. \textbf{c}, Cross-sectional schematic of the device structure. \textbf{d}, Optical transmission spectrum (red curve) and fluorescence spectrum (blue curve) of the device. Inset: close-up of the resonance studied in this work. The shaded area indicates the 202-GHz inhomogeneous broadening of the erbium ions. \textbf{e}, Power-dependent transmission spectra. \textbf{f}, The loaded quality factor ($Q_{\text{loaded}}$) and extinction ratio (ER) extracted from Fano fits as a function of the input power, revealing the saturation dynamics of the erbium ensemble embedded in the cavity. Error bars represent 95\% confidence intervals propagated in quadrature from Poisson shot noise, attenuation calibration errors, and Fano-fit parameter uncertainties. See text for details.}
\label{fig1}
\end{figure*}

Fig.~\ref{fig1}a illustrates the conceptual layout of our $^{167}\text{Er}^{3+}$-doped TFLN device: a bus waveguide evanescently coupled to a racetrack microring resonator with a 2759-$\mu$m perimeter (Fig.~\ref{fig1}b). The device is fabricated on a TFLN wafer, with a top LN layer doped with isotopically purified $^{167}\text{Er}^{3+}$ ions (95.6\% nominal isotopic purity) during bulk crystal growth. The cross-section (Fig.~\ref{fig1}c) consists of a 1.2-$\mu$m-wide ridge waveguide on a 300-nm-thick X-cut LN film with symmetric SiO$_2$ cladding (mode field diameter $\sim$0.57~$\mu$m), supporting a TE mode configured to access LN's $r_{33}$ electro-optic coefficient; gold electrodes alongside the straight sections of the resonator enable direct cavity tuning. Two independent chips with $^{167}\text{Er}^{3+}$ doping levels of 500~ppm (DEV1) and 50~ppm (DEV2) were fabricated to vary the strength of the light--ion interaction and the resulting ensemble absorption (Fabrication details in Supplementary Section S1).

The chips are independently mounted on a 260~mK stage of a dilution refrigerator and accessed through actively aligned cryogenic fiber arrays (single-facet loss $\sim$7.8~dB). The optical transmission of DEV1 (Fig.~\ref{fig1}d, red) shows a strongly over-coupled resonance near 1532~nm with a loaded quality factor $Q_{\text{loaded}} = 1.78 \times 10^5$ (inset), which corresponds to an external coupling rate from bus waveguide to resonator $\kappa_{\text{ext}}/2\pi = 991$~MHz and an intrinsic optical loss of the resonator $\kappa_{\text{loss}}/2\pi = 119$~MHz. Sweeping the pump across the $^{4}I_{15/2}\leftrightarrow{}^{4}I_{13/2}$ transition while electro-optically tracking the cavity yields the inhomogeneous absorption profile (shaded area in Fig.~\ref{fig1}d), centered at 1532.13~nm with a 202-GHz full width---a broad spectral resource for telecom memories. The fluorescence spectrum (blue curve in Fig.~\ref{fig1}d) peaks when the pump is on resonance with a cavity mode, consistent with most of the ions contributing to the measured signal being located inside the microring, where the intracavity field is strongly enhanced. We further verify this cavity--ion coupling by comparing the fluorescence lifetimes of DEV1 and DEV2 on- and off-resonance. A clear cavity-induced shortening of the excited-state lifetime via the Purcell effect is observed (Supplementary Section S2), which helps efficient spectral hole burning.

To quantify the cavity--ion interaction, we sweep a continuous-wave (CW) probe across the resonance closest to the absorption center with on-chip optical power from sub-picowatt up to above a microwatt, and record the attenuated output with an 80\%-efficient superconducting nanowire single-photon detector (SNSPD) (Fig.~\ref{fig1}e). The quantitative evolution of the cavity--ion system is summarized in Fig.~\ref{fig1}f, which presents $Q_{\text{loaded}}$ and the extinction ratio (ER) extracted from Fano fits of the resonance across the full measured power range. As the erbium ions saturate with increasing input power, $\kappa_{\text{ions}}$ asymptotically approaches zero, resulting in a monotonic rise of $Q_{\text{loaded}}$ from $6.8 \times 10^4$ to $1.5 \times 10^5$. By comparing the total loss rate at the lowest input power, $(\kappa_{\text{ext}} + \kappa_{\text{loss}} + \kappa_{\text{ions}})/2\pi$, with the ion-free cavity loss rate, $(\kappa_{\text{ext}} + \kappa_{\text{loss}})/2\pi$ (determined from Fig.~\ref{fig1}d), we deduce the additional loss introduced by the erbium ensemble to be about $\kappa_{\text{ions}}/2\pi = 1778$~MHz. The evolution of ER can be understood physically as follows: the cavity is deliberately designed to be strongly over-coupled in the absence of ion absorption ($\kappa_{\text{ext}} > \kappa_{\text{loss}}$), so that at low powers the large ion absorption pulls the system closer to critical coupling ($\kappa_{\text{ext}} \approx \kappa_{\text{loss}} + \kappa_{\text{ions}}$) and yields a high ER, while at high powers the ions cease to contribute to the cavity loss and the system returns to its over-coupled regime. Control measurements on DEV2 and at absorption-free wavelengths, together with a self-consistency model, confirm that this evolution is ion-induced (Supplementary Sections S2 and S3).

\subsection*{Quantum Storage Using Atomic Frequency Comb}

A persistent, high-contrast atomic frequency comb (AFC)~\cite{afzeliusMultimodeQuantumMemory2009} requires a long-lived population reservoir to shelve the ions optically pumped away during spectral tailoring of the AFC. Our isotopically purified $^{167}\text{Er}^{3+}$ sample fulfills this requirement through the nuclear-hyperfine shelving states of the $^{167}\text{Er}^{3+}$ ground manifold. Note that naturally abundant Er:LN lacks such long-lived states and is limited by rapid hole refilling~\cite{zhangTelecombandintegratedMultimodePhotonic2023,baryaUltraHighQTunable2025a}.

\begin{figure*}[t]
\centering
\includegraphics[width=\textwidth]{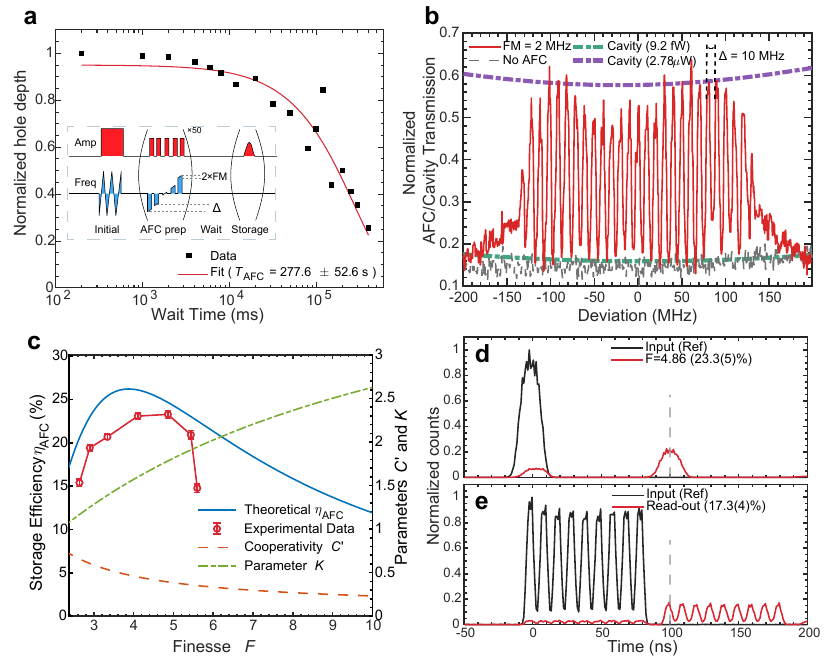}
\caption{\textbf{Persistent atomic frequency combs and quantum storage efficiency.} \textbf{a}, Normalized hole depth as a function of wait time. The single-exponential fit yields an AFC lifetime of $277.6(52.6)$~s. Inset: Experimental pulse sequence. \textbf{b}, Normalized transmission spectrum of AFC (red solid line) prepared at a frequency modulation (FM) amplitude of 2~MHz. An unpumped probe scan (grey dashed line) confirms that the readout does not perturb the AFC. \textbf{c}, Storage efficiency (red circles), theoretical efficiency curve (blue solid line), coupling parameter $K$ (green dashed line), and effective cooperativity $C'$ as functions of the comb finesse $F$. The theoretical curve assumes a preparation efficiency of $\eta_{\text{spectral}} = 0.95$, close to that obtained from \textbf{b}. \textbf{d}, Time-resolved histogram of the retrieved signal for a 100-ns storage time using 15-ns input pulses, yielding an optimal efficiency of $23.3(5)\%$. \textbf{e}, Temporally multiplexed quantum memory. Nine consecutive 5-ns pulses are stored and retrieved after 100~ns with a collective efficiency of $17.3(4)\%$.}
\label{fig2}
\end{figure*}

The experimental sequence is illustrated in the inset of Fig.~\ref{fig2}a. We first initialize the ion population by repeatedly sweeping a CW laser across 1528--1536~nm. The AFC is then burned by a Pound-Drever-Hall (PDH)-stabilized pump laser---locked to an external Fabry--P\'erot (F-P) cavity---whose frequency is shifted by a single-sideband (SSB) modulator to address each designated comb tooth. At each tooth, the SSB applies a linear triangular frequency chirp with a frequency-modulation amplitude denoted by FM, which sets the width of the burned spectral hole and hence that of the resulting comb tooth. Each tooth is burned in 10-ms on/off cycles, and the full burning sequence is repeated 50 times. A 200-ms wait is then inserted before either a photon-storage experiment or a spectral measurement, allowing residual fluorescence to decay (full setup shown in Supplementary Section S5).

To measure the AFC lifetime, we vary the waiting interval and monitor the hole depth as a function of delay. A single-exponential fit to the resulting decay curve (Fig.~\ref{fig2}a) gives $T_{\mathrm{AFC}} = 277.6(52.6)$~s, supporting low background absorption, and hence high-efficiency storage. This substantially exceeds previous results on naturally abundant erbium samples, where the AFC decay required a bi-exponential fit with components of $0.55$~s and $32.75$~s~\cite{zhangTelecombandintegratedMultimodePhotonic2023}; in that case the short component acts as a leakage channel during optical pumping and limits the achievable hole-burning depth. We further measure the optical coherence of the ions in our device. A post-etching oxygen anneal mitigates the dry-etch--induced lattice damage, yielding a Hahn-echo coherence time $T_{2} = 93.0(4.8)~\mu$s (Supplementary Section S6).

A second factor limiting the contrast of the AFC is the superhyperfine coupling of $^{167}\text{Er}^{3+}$ to host $^{93}\text{Nb}$ and $^7\text{Li}$ nuclei, which generates field-dependent side-holes (Supplementary Section S7). We tune the magnetic field so that the side-holes coincide with the troughs rather than falling on the teeth; for $\Delta = 10$~MHz (100-ns storage), the optimal field is 1.855~T, which shifts the Nb and Li side-holes to approximately 20~MHz and 30~MHz. Under these conditions, Fig.~\ref{fig2}b displays the normalized transmission spectrum of a 21-tooth AFC prepared with $\text{FM} = 2$~MHz. Evaluated using a weak 6.5-fW probe, the comb profile exhibits sharp modulation between the unpumped baseline (green dashed line) and the fully saturated cavity limit (purple dashed line)---corresponding to the transmission spectra at the lowest (green square) and highest (purple square) probe powers identified in Fig.~\ref{fig1}f---indicating near-unity preparation efficiency ($\eta_{\text{spectral}} \approx 0.95$). Control measurements with the same weak probe but without AFC preparation (grey dashed line) confirm that the readout process does not perturb the prepared AFC.

Embedding the erbium ensemble inside a microring substantially enhances the effective light--matter interaction, compensating for the limited single-pass optical depth of the erbium ensemble. In this cavity-enhanced setting, perfect absorption is reached when the external coupling exactly balances the total internal loss rate, fulfilling the impedance-matching condition~\cite{afzeliusImpedancematchedCavityQuantum2010,moiseevEfficientMultimodeQuantum2010}. While this condition has been met with rare-earth ion ensembles in bulk crystals~\cite{sabooniEfficientQuantumMemory2013,jobezCavityenhancedStorageOptical2014,davidsonImprovedLightmatterInteraction2020,durantiEfficientCavityassistedStorage2024} and in integrated Nd$^{3+}$~\cite{zhongNanophotonicRareearthQuantum2017a} and Eu$^{3+}$ ion~\cite{mengEfficientIntegratedQuantum2026} devices, prior on-chip cavity-enhanced erbium memories at telecom wavelengths have been constrained by relatively high intrinsic cavity losses and have not reached impedance matching~\cite{craiciuNanophotonicQuantumStorage2019,craiciuMultifunctionalOnchipStorage2021}, limiting their on-chip storage efficiency. The storage efficiency of the cavity-enhanced AFC memory is~\cite{afzeliusImpedancematchedCavityQuantum2010,moiseevEfficientMultimodeQuantum2010,zhongNanophotonicRareearthQuantum2017a}:
\begin{equation}
\eta_{\text{AFC}} = \left[ \frac{1}{F(1/\eta_{\text{spectral}} - 1) + 1} \frac{\kappa_{\text{ext}}}{\kappa_{\text{total}}} \frac{4C'}{(1+C')^2} \right]^2 \eta_d,
\end{equation}
with $\eta_d = \exp\!\left(-\pi^2/(2\ln 2\cdot F^2)\right)$ the intrinsic phase-decoherence factor, $\kappa_{\text{total}} = \kappa_{\text{ext}} + \kappa_{\text{loss}}$, and the effective cooperativity $C' = \left( \eta_{\text{spectral}}/F + (1 - \eta_{\text{spectral}}) \right) C$ that couples the comb finesse $F$ and the preparation efficiency $\eta_{\text{spectral}}$ through the bare cooperativity $C = \kappa_{\text{ions}}/\kappa_{\text{total}}$.

Guided by this framework, we tune $F$ via the FM amplitude (Supplementary Section S8) and track the coupling parameter $K = \kappa_{\text{ext}}/(\kappa_{\text{loss}} + \kappa_{\text{ions}}^{\text{eff}})$, where $\kappa_{\text{ions}}^{\text{eff}}$ includes the residual absorption from imperfect preparation. As shown in Fig.~\ref{fig2}c, the optimum efficiency $23.3(5)\%$ occurs at $F \approx 4.86$, away from impedance matching ($K=1$), owing to the unavoidable $\kappa_{\text{loss}}$; the experimental efficiencies (red circles) closely follow the theoretical model (blue line) with $\eta_{\text{spectral}}=0.95$. The corresponding single-mode retrieval trace at 100-ns delay is shown in Fig.~\ref{fig2}d, where the input consists of weak coherent pulses with a 15-ns Gaussian profile and an average photon number of $\mu = 0.163(1)$ per pulse; integrating the retrieved counts over a 30-ns window and normalizing to the off-resonance input reference yields a $23.3(5)\%$ storage efficiency, which is comparable to the highest storage efficiency reported so far for quantum memories operating in the telecom band~\cite{stuartInitializationProtocolEfficient2021}. Exploiting the broad inhomogeneous bandwidth of the ensemble, we demonstrate robust temporal multiplexing on the same device (Fig.~\ref{fig2}e): a 41-tooth AFC over 400-MHz bandwidth stores 9 sequential 5-ns modes within 100~ns at $17.3(4)\%$, and up to 18 modes within 200~ns at $6.0(2)\%$. Additional data at other storage times and temporal modes are provided in Supplementary Sections S9 and S10.

\subsection*{High Speed Routing of Quantum Memory}

Building on this cavity--ion interface, a defining advantage of our monolithic TFLN platform is the strong Pockels effect~\cite{boesLithiumNiobatePhotonics2023}, which enables high-speed electro-optic control of the cavity resonance. Prior to quantum storage, we characterize the intrinsic electro-optic response of the microring and obtain an EO tuning efficiency of about 554~MHz/V together with a broad modulation bandwidth that is consistent with our finite-element simulations (Supplementary Sections S11 and S12). For the dynamic routing experiments in this section, we employ DEV1 to demonstrate electro-optically controlled frequency-selective storage and retrieval in the same cavity--ion platform.

\begin{figure*}[t]
\centering
\includegraphics[width=\textwidth]{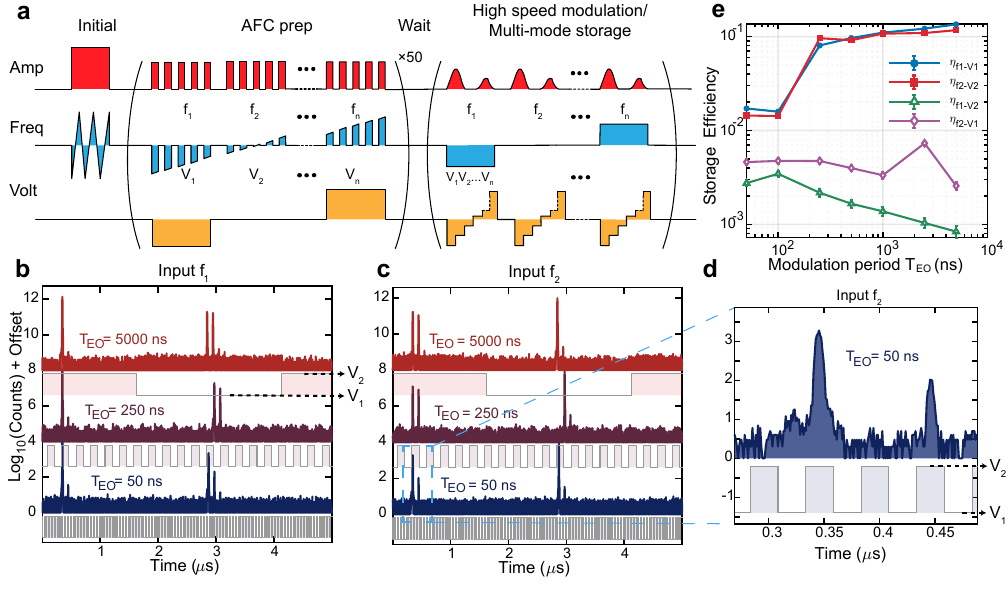}
\caption{\textbf{High-speed routing of the quantum memory.} \textbf{a}, Experimental pulse sequence. Distinct DC voltages ($V_n$) shift the cavity resonance to target channels to prepare independent AFCs ($f_n$). During the storage phase, high-speed voltage modulation dynamically routes the retrieved photons. \textbf{b}, Time-resolved histograms for input photons fixed at $f_1$ under square-wave voltage modulations. Echoes are retrieved predominantly during $V_1$ segments, while pulses falling under $V_2$ are suppressed. \textbf{c}, Same measurement as in \textbf{b}, but with the input photon frequency fixed at $f_2$, demonstrating frequency-selective dynamic routing complementary to \textbf{b}. \textbf{d}, Zoomed view of the $f_2$ input data at $T_{\mathrm{EO}} = 50$~ns, illustrating that the echo is recovered after the cavity resonance is shifted away and restored within one storage period. \textbf{e}, Matched storage efficiencies ($\eta_{f_1-V_1}$, $\eta_{f_2-V_2}$) and unwanted retrieval efficiency ($\eta_{f_1-V_2}$, $\eta_{f_2-V_1}$) as a function of the modulation period $T_{\mathrm{EO}}$. See text for details.}
\label{fig3}
\end{figure*}

The experimental pulse sequence for frequency-selective operation is outlined in Fig.~\ref{fig3}a. During the AFC preparation phase, we apply distinct DC voltages to the microring electrodes, shifting its resonance to two target frequency channels while independent AFCs are sequentially burned. We probe dynamic frequency addressing by applying high-speed electro-optic modulation within the memory window. Two predetermined frequency channels $f_1$ and $f_2$ are addressed by two voltage configurations $V_1$ and $V_2$, defined such that $V_1=-1.5$~V tunes the microring resonance to $f_1=6758$~MHz and $V_2=+1.5$~V tunes the resonance to $f_2=8508$~MHz. At 1.855~T and 100-ns storage, we drive the electrodes with a square wave toggling between $V_1$ and $V_2$ at 50\% duty cycle, and align two 5-ns Gaussian input pulses per period to the centers of the two half-cycles. The stacked histograms (Fig.~\ref{fig3}b,c) show frequency-selective retrieval: when the input is at $f_1$, the echo is obtained during the $V_1$-aligned configuration, corresponding to the matched storage efficiency $\eta_{f_1-V_1}$, while the signal measured under the mismatched $V_2$ configuration is denoted as the unwanted retrieval efficiency $\eta_{f_1-V_2}$; when the input is at $f_2$, the echo is obtained during the $V_2$-aligned configuration, corresponding to the matched storage efficiency $\eta_{f_2-V_2}$, while the signal measured under the mismatched $V_1$ configuration is denoted as the unwanted retrieval efficiency $\eta_{f_2-V_1}$. A noticeable residual crosstalk is observed, which we attribute to the broader cavity linewidth and stronger absorption associated with the higher erbium doping concentration for DEV1.

To explore the high-speed limit of this active storage routing, we progressively reduce the electro-optic modulation period $T_{\mathrm{EO}}$ from 5000~ns down to 50~ns. At $T_{\mathrm{EO}}=50$~ns, as shown in Fig.~\ref{fig3}d, we first store the input at $f_2$ into the cavity--ion device under the $V_2$ configuration with cavity enhancement, then dynamically shift the cavity resonance between the $V_1$ and $V_2$ configurations three times, each with 25-ns duration. After 75~ns, we tune back to $V_2$ and read out the echoes at about 100~ns, confirming that the full store--shift--echo cycle completes within a single storage period. It is intriguing that the coherence of the erbium ensemble remains sufficiently preserved to enable echo retrieval during the reconfiguration of the cavity resonance, which is unique to our TFLN platform and provides a new tool for controlling quantum memories.

Fig.~\ref{fig3}e shows that significant echoes are retrieved across two orders of magnitude in modulation speed. Based on the efficiencies defined above, we define the corresponding inter-channel crosstalk as
$\chi_{f_1-V_2}=\eta_{f_1-V_2}/\eta_{f_1-V_1}$ and
$\chi_{f_2-V_1}=\eta_{f_2-V_1}/\eta_{f_2-V_2}$, which remain on the order of $10^{-1}$ over the measured modulation range. The dependence of residual crosstalk on the applied detuning voltage is further characterized in Supplementary Section S13. To suppress this residual crosstalk, we additionally investigate DEV2, whose lower doping level and weaker external coupling together yield a higher loaded quality factor of $3.6 \times 10^5$ and a correspondingly narrower cavity linewidth, suited to minimizing inter-channel crosstalk. We verify static frequency-selective operation over frequency shifts up to 2.69~GHz, with negligible excess loss and inter-channel crosstalk suppressed below $10^{-2}$ in DEV2 (Supplementary Section S14). Further measurements of DEV2, including storage efficiency at different storage times and dynamic routing performance, are provided in Supplementary Sections S15 and S16.

\subsection*{Storage of Photonic Entanglement}

\begin{figure*}[t]
\centering
\includegraphics[width=\textwidth]{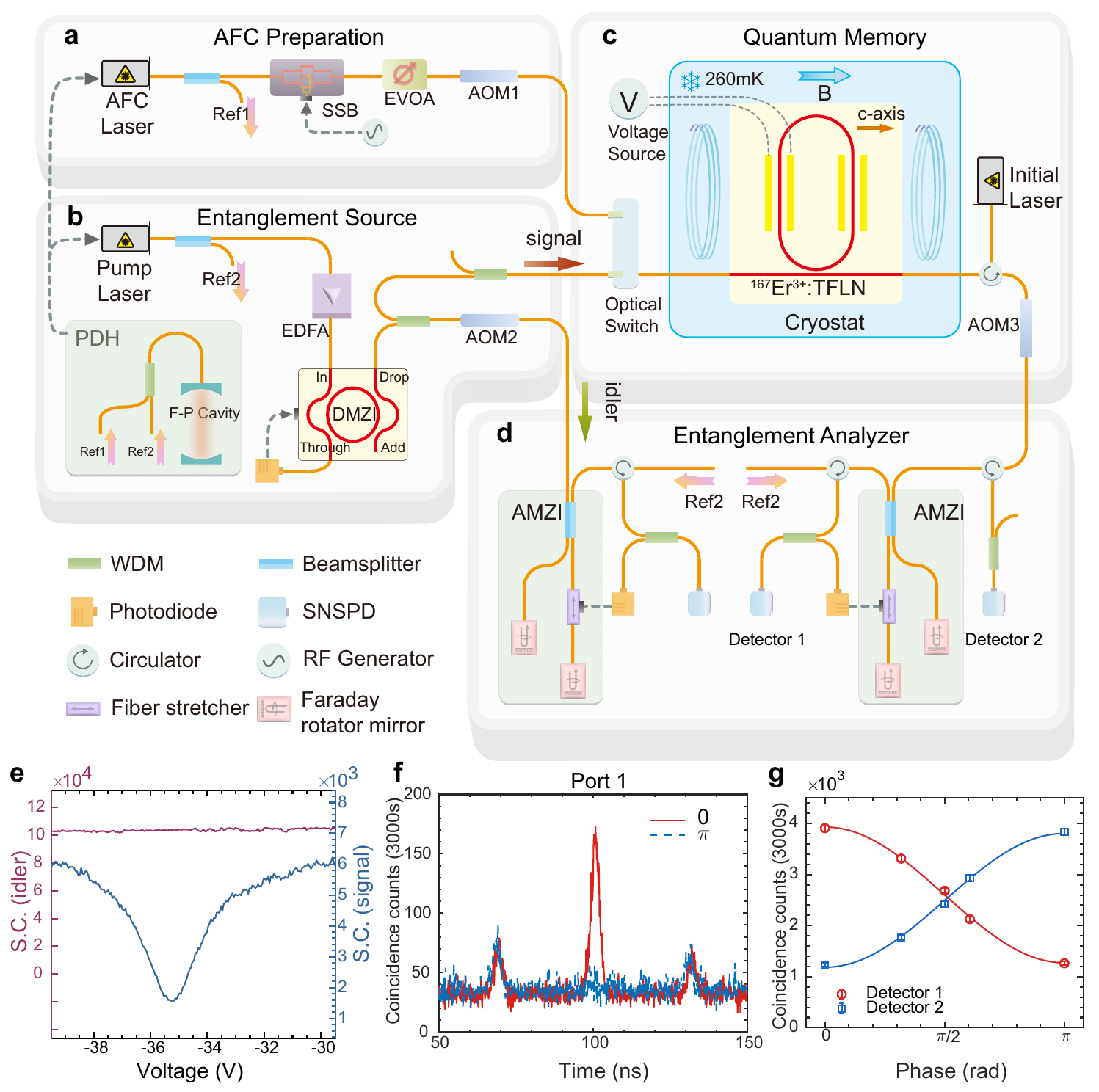}
\caption{\textbf{Storage of time-energy entangled telecom photons.} \textbf{a}, AFC preparation module. \textbf{b}, Integrated entangled photon-pair source based on a silicon nitride dual Mach--Zehnder interferometer (DMZI) microring resonator. \textbf{c}, The $^{167}\text{Er}^{3+}$:TFLN quantum memory. \textbf{d}, Entanglement analyzer consisting of two unbalanced asymmetric Mach--Zehnder interferometers (AMZIs). \textbf{e}, Frequency alignment between the memory cavity and the input signal photon. The clear dip in the signal counts (S.C.) indicates that the cavity resonance has been electro-optically tuned into resonance with the fixed signal wavelength, while the idler counts remain constant as expected. \textbf{f}, Time-resolved coincidence counts between the idler and the retrieved signal photon (after 100-ns storage) at AMZI phases of 0 and $\pi$. \textbf{g}, Two-photon interference fringes as a function of the AMZI phase. See text for details.}
\label{fig4}
\end{figure*}

To unambiguously verify the quantum nature of the memory, we store time-energy-entangled telecom photons generated by a triple-critical-coupled silicon-nitride dual Mach--Zehnder interferometer microring (DMZI-R, Fig.~\ref{fig4}b), as developed in our previous work~\cite{jiangQuantumStorageEntangled2023,anQuantumTeleportationTelecom2025}. CW pumping at 1538.92~nm produces signal (1531.89~nm) and idler (1546.02~nm) photons by spontaneous four-wave mixing with $\sim$200-MHz bandwidth, matched to that of the prepared AFC (details of the entangled photon source are shown in Supplementary Section S17). After spectral separation by wavelength-division multiplexers, the idler is routed to the entanglement analyzer while the signal enters the quantum memory (Fig.~\ref{fig4}a,c). Coincidence events between the heralding idler and the retrieved signal are recorded by a time-to-digital converter. Because the DMZI-R operates at a fixed pump wavelength stabilized by an external F-P cavity, the signal wavelength is fixed as well. To align the memory to it, we harness the electro-optic tunability of the TFLN microring (Fig.~\ref{fig4}e): sweeping the DC voltage applied to DEV1 tunes the cavity resonance across the signal frequency, which is identified by a dip in the transmitted signal single counts that signifies on-resonance alignment with the cavity. The AFC preparation laser is then tuned to this resonance center. Under optimized alignment, the on-chip storage efficiency for the entangled signal photons reaches $8.6(5)\%$. The same electro-optic control also provides a natural knob for long-term stabilization: a feedback loop periodically (e.g., every 1~hour) readjusts the bias voltage based on the signal single counts, counteracting slow resonance drift and maintaining stable storage efficiency throughout the 10-hour acquisition (Supplementary Section S18).

With the memory frequency-aligned, the signal photons are stored for 100~ns, heralded by the idler detections. The cross-correlation between idler and retrieved signal is $g_{si}^{(2)}(0) = 4.54(30)$, well above the classical thermal-light bound of 2, confirming that non-classical correlations are preserved by the cavity-enhanced storage. To verify the existence of entanglement, both photons are then analyzed by a Franson interferometer~\cite{fransonBellInequalityPosition1989} consisting of two actively stabilized, unbalanced asymmetric Mach--Zehnder interferometers (AMZIs, Fig.~\ref{fig4}d); both the interferometer phases and the reference laser's frequency are locked~\cite{anQuantumTeleportationTelecom2025}. Phase-resolved coincidence measurements of the central temporal peak---corresponding to indistinguishable paths where both photons traverse either the short or the long arms---yield interference fringes (Fig.~\ref{fig4}f,g), with visibilities at the two signal output ports of $\mathcal{V}_1 = 51.17(1.19)\%$ and $\mathcal{V}_2 = 51.30(1.21)\%$. To rigorously verify entanglement, we obtain the expectation value of the entanglement witness $\langle W \rangle = \frac{1}{g_{si}^{(2)}(0) + 2} - \frac{\mathcal{V}}{2}$~\cite{jiangQuantumStorageEntangled2023}, where $\mathcal{V}$ is the mean of $\mathcal{V}_1$ and $\mathcal{V}_2$. We obtain $\langle W \rangle = -0.1037(92)$, violating the separable-state bound ($W \geq 0$) by more than 11 standard deviations. This violation of the classical bound is obtained directly from the raw data, without any subtraction of background noise or accidental coincidences.

\subsection*{Conclusion and Outlook}

In summary, we have realized an integrated erbium-doped thin-film lithium niobate quantum memory that unites persistent AFC preparation, cavity impedance matching, and fast electro-optic spectral control. The platform supports an on-chip efficiency of $23.3(5)\%$ for 100-ns storage, temporal multiplexing up to 18 modes for 200-ns storage, frequency-selective operation over a 2.69-GHz range, electro-optic routing at rates up to 20~MHz, and the storage and retrieval of time-energy-entangled telecom photons.

Extending the memory time is the next key challenge. Longer storage time requires narrower comb teeth, ultimately limited by how narrow a persistent hole can be burned and preserved against spectral diffusion~\cite{thielOpticalDecoherencePersistent2010,askaraniPersistentAtomicFrequency2020,zhangTelecombandintegratedMultimodePhotonic2023}. Systematically optimized post-etch annealing~\cite{thielRareearthdopedMaterialsApplications2011,lutzEffectsMechanicalProcessing2017,duttaIntegratedPhotonicPlatform2020}, ultra-low temperatures, and stronger magnetic fields~\cite{zhongOpticallyAddressableNuclear2015,rancicCoherenceTimeSecond2018,yuFrequencyTunableCavityEnhanced2023} are all effective routes. Combined with reduced propagation losses---state-of-the-art TFLN microrings reach intrinsic $Q > 10^7$~\cite{zhuTwentynineMillionIntrinsic2024b}---and deeper hole burning, storage efficiencies exceeding 70\% could potentially be reached.

More broadly, the significance of this platform lies in uniting memory, frequency addressing, and high-speed control in one material system~\cite{zhouPhotonicIntegratedQuantum2023}: the 202-GHz inhomogeneous bandwidth supplies a spectral resource that EO tuning makes electrically addressable on chip, opening routes to dense spectral multiplexing~\cite{sinclairSpectralMultiplexingScalable2014} and random-access architectures~\cite{puExperimentalRealizationMultiplexed2017,tellerQuantumStorageQubits2025,ouMultichannelHighDimensional2025} in the telecom band. Our work has combined an integrated photon-pair source with an integrated quantum memory, outlining a practical path toward fully integrated, programmable quantum repeater networks~\cite{pelucchiPotentialGlobalOutlook2021,zhouPhotonicIntegratedQuantum2023}.

\section*{Acknowledgments}
\paragraph*{Funding:} This research was supported by HFNL Self-Deployed Project (ZB2026020200), the Natural Science Foundation of Jiangsu Province (Grants Nos. BK20240006, BK20233001), Quantum Science and Technology-National Science and Technology Major Project (Grants Nos. 2021ZD0300700 and 2021ZD0301500), Fundamental and Interdisciplinary Disciplines Breakthrough Plan of the Ministry of Education of China (JYB2025XDXM112), and Nanjing University-China Mobile Communications Group Co., Ltd. Joint Institute.
\section*{Author contributions:} C.Y., H.G., Y.-Y.A., Q.H., C.L., and Z.J. constructed the experimental setup. C.Y., Y.-Y.A., and Z.J. carried out the measurements. C.Y., Q.H., and C.L. processed and analyzed the experimental data. C.Y. and H.G. prepared the figures. C.Y. designed the wafer sample and chip. C.Y. and X.-S.M. wrote the manuscript with input from all the authors. Y.-Q.L., S.Z., and X.-S.M. supervised the project.
\section*{Competing interests:} The authors declare no competing interests.
\section*{Data availability:} The data that support the plots within this paper and other findings of this study are available from the corresponding author upon reasonable request.

\newpage
\onecolumngrid
\def\scititle{Programmable cavity-enhanced telecom quantum memory in thin-film lithium niobate}

\renewcommand{\thefigure}{S\arabic{figure}}
\renewcommand{\thetable}{S\arabic{table}}
\renewcommand{\theequation}{S\arabic{equation}}
\setcounter{figure}{0}
\setcounter{table}{0}
\setcounter{equation}{0}
\setcounter{secnumdepth}{0}

\begin{center}
\section*{Supplementary Materials for \scititle}
Chengdong~Yang, Hanwen~Guo, Yu-Yang~An, Qian~He, Chi~Lu, Ziheng~Jiang, Yan-Qing~Lu, Shining~Zhu, Xiao-Song~Ma$^{\ast}$\\
\small$^\ast$Corresponding author. Email: Xiaosong.Ma@nju.edu.cn
\end{center}

\subsubsection{S1. Device Fabrication and Packaging}

The bulk lithium niobate crystal doped with isotopically purified $^{167}\text{Er}^{3+}$ was grown using the Czochralski (CZ) method by Shanghai Chenghan Optical Precision Machinery Co., Ltd. The thin-film lithium niobate (TFLN) wafer was then prepared via a commercial smart-cut process by NanoLN. Specifically, helium ions were implanted into the doped bulk crystal, which was subsequently bonded to a silicon substrate carrying a 4.7-$\mu$m-thick pre-oxidized silicon dioxide (SiO$_2$) buried-oxide (BOX) layer. Following a precision heat treatment to exfoliate the film, subsequent annealing and chemical-mechanical polishing (CMP) yielded a high-quality, 300-nm-thick X-cut TFLN layer on the insulator substrate.

\begin{figure}[htbp]
\centering
\includegraphics[width=\linewidth]{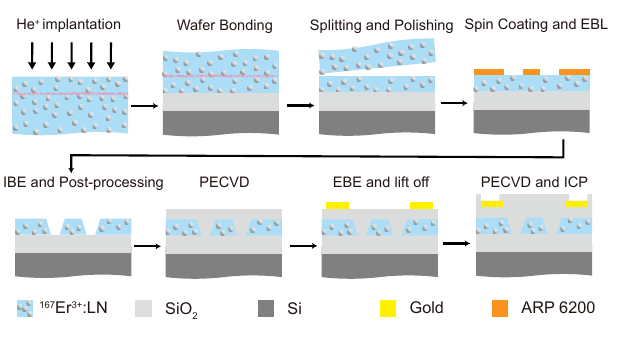}
\caption{\label{figS1} \textbf{Fabrication flow of the $^{167}\text{Er}^{3+}$-doped TFLN quantum memory.} The process begins with He$^+$ implantation of the bulk-grown erbium-doped crystal, followed by wafer bonding, smart-cut splitting, and polishing to produce the TFLN substrate. Photonic structures are then defined by EBL patterning and transferred by IBE etching with subsequent post-etch processing. Finally, a SiO$_2$ cladding and Cr/Au electrodes are integrated by PECVD and electron-beam evaporation. Materials are identified in the legend. See text for details.}
\end{figure}

The subsequent device fabrication was carried out by IOPTEE. Photonic waveguide and microring structures were defined by electron-beam lithography (EBL) with a high-resolution resist ARP 6200, and the patterns were transferred into the TFLN layer by argon-based ion-beam etching (IBE). The 300-nm TFLN film was fully etched to the buried oxide to form ridge waveguides with a nominal width of 1.2~$\mu$m. The racetrack microring has a perimeter of 2759~$\mu$m, corresponding to a free spectral range of 50~GHz, and uses a continuously varying Euler bend in the coupling region with a minimum bending radius of approximately 40.1~$\mu$m and a 0.5-$\mu$m gap to the bus waveguide. To mitigate scattering losses, a standard RCA cleaning was applied to remove etching redeposits. For the 500-ppm device (DEV1), an additional thermal annealing step was performed in an oxygen-rich environment at 500$^{\circ}$C for 2~hours; this step repairs the crystalline lattice damage and oxygen vacancies induced by dry etching, thereby preserving the optical coherence time of the erbium ions (see Supplementary Supplementary Section S6 for detailed coherence measurements).

To prevent absorption losses caused by the metal electrodes and to improve step coverage for subsequent metallization, an 800-nm-thick SiO$_2$ buffer layer was deposited by plasma-enhanced chemical vapor deposition (PECVD). The metal electrodes were then patterned by ultraviolet (UV) lithography with a 2-$\mu$m gap from the waveguide edge. A 10-nm Cr adhesion layer and a 400-nm Au layer were deposited by electron-beam evaporation, followed by a standard lift-off process. An additional 3-$\mu$m SiO$_2$ layer was then deposited by PECVD on top of the buffer, yielding a total upper cladding thickness of 3.8~$\mu$m. This symmetric index profile ensures spatial mode-field matching with the high-numerical-aperture fiber (HNAF) arrays for edge coupling. Finally, via windows were opened through the top cladding by UV lithography and inductively coupled plasma (ICP) etching to expose the electrode pads.

The fabricated wafer was diced into individual chips, and the end facets were polished to optical-grade quality to construct high-efficiency edge couplers. The chip was mounted on a customized copper sample holder using silver epoxy to ensure good thermal anchoring at millikelvin temperatures, and the on-chip electrode pads were wire-bonded to a printed circuit board (PCB) for high-speed microwave delivery. During operation, the sample holder was mounted on a cryogenic XYZ nanopositioning stage to maintain precise optical alignment with the HNAF arrays.

\subsubsection{S2. Fluorescence Lifetime and Purcell Enhancement of $^{167}\text{Er}^{3+}$ in TFLN}

\begin{figure}[htbp]
\centering
\includegraphics[width=\linewidth]{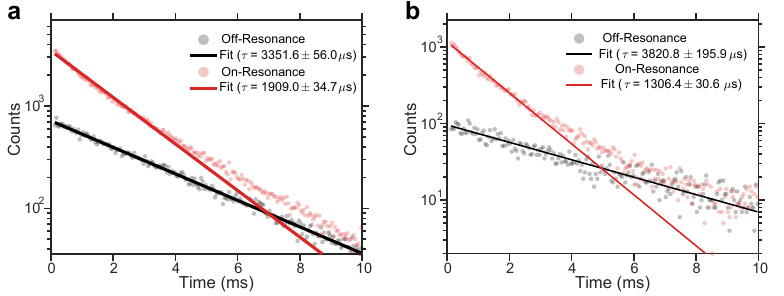}
\caption{\label{figS2} \textbf{Fluorescence lifetime and Purcell enhancement of $^{167}\text{Er}^{3+}$ in TFLN.} Time-resolved fluorescence decay for \textbf{a}, DEV1 and \textbf{b}, DEV2 TFLN devices. Grey and red dots represent off-resonance (uncoupled) and on-resonance (cavity-coupled) measurements, with corresponding single-exponential fits in black and red solid lines, respectively. The extracted Purcell factors are $F_p = 1.75(4)$ and $2.92(17)$, respectively.}
\end{figure}

We quantify the cavity Purcell enhancement by measuring the time-resolved fluorescence decay of the $^{167}\text{Er}^{3+}$ ions. Measurements are performed for both DEV1 and DEV2 TFLN devices under ``uncoupled'' (off-resonance, probing bus waveguide only) and ``cavity-coupled'' (on-resonance) conditions. As shown in Fig.~\ref{figS2}, the intrinsic excited-state lifetimes for the uncoupled ions are $3.35(6)$~ms (DEV1) and $3.82(20)$~ms (DEV2). When resonantly coupled to the microring, the Purcell effect accelerates the spontaneous emission, reducing the lifetimes to $1.91(3)$ ms and $1.31(3)$ ms, respectively. This yields Purcell factors ($F_p = \tau_{wg} / \tau_{cavity}$) of $1.75(4)$ for DEV1 and $2.92(17)$ for DEV2. The higher enhancement in DEV2 stems from its larger loaded $Q$ factor.

\subsubsection{S3. Control Measurements for Power-Dependent Cavity Dynamics}

\begin{figure}[htbp]
\centering
\includegraphics[width=\linewidth]{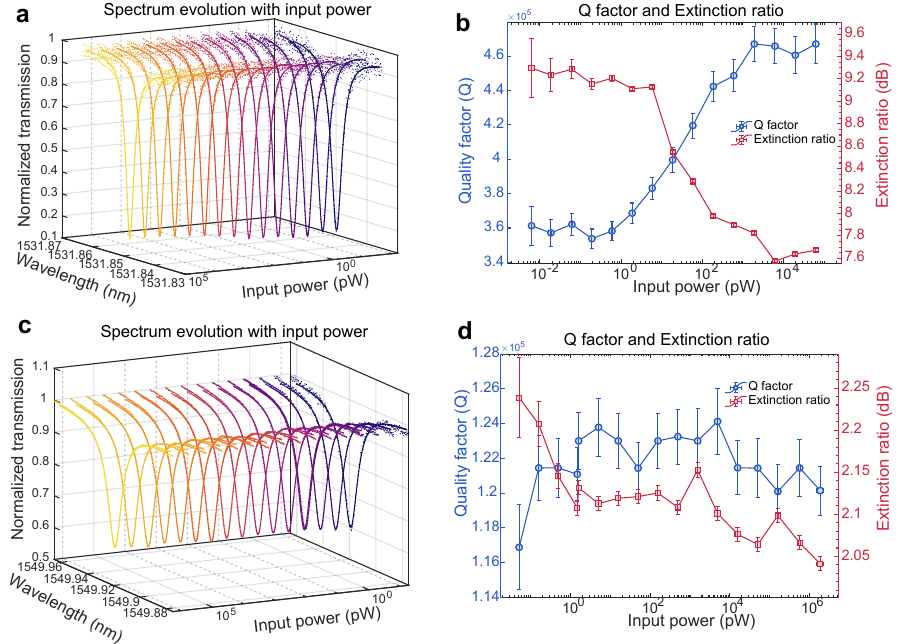}
\caption{\label{figS3} \textbf{Power-dependent cavity dynamics for DEV2 and at an absorption-free wavelength.} \textbf{a,b}, Power-dependent transmission spectra around the resonance at 1531.85~nm and extracted $Q$ and ER for DEV2. \textbf{c,d}, Control measurements of DEV1 at 1549.92~nm, a wavelength outside the erbium absorption band, showing nearly constant $Q$ and ER across all power levels and confirming the absence of parasitic nonlinearities.}
\end{figure}

To evaluate the interaction between the microring and the erbium ensemble, we performed power-dependent transmission measurements using a tunable laser (Santec TSL-570) with variable optical attenuators (VOAs) to regulate on-chip power and protect the detector. For the 50-ppm doped device (Fig.~\ref{figS3}a,b), as on-chip power decreases from the saturation to the single-photon level, the loaded quality factor $Q_{\text{loaded}}$ decreases from $4.6 \times 10^5$ to $3.6 \times 10^5$, while the extinction ratio (ER) increases from 7.5 dB to 9.0 dB. Fitting these dynamics yields $\kappa_{\text{loss}}/2\pi \approx 123$ MHz, $\kappa_{\text{ext}}/2\pi \approx 302$ MHz, and $\kappa_{\text{ions}}/2\pi \approx 117$ MHz. As expected from the lower doping concentration, this ion coupling rate is significantly weaker than the $\kappa_{\text{ions}}/2\pi = 1778$ MHz observed in the heavily doped DEV1 discussed in the main text, resulting in less pronounced power-dependent variations. To confirm the origin of these dynamics, control measurements were conducted at 1549.92~nm, a wavelength outside the erbium absorption band (Fig.~\ref{figS3}c,d). The $Q$ and ER remain almost constant across all power levels, verifying that the modulation near 1532~nm arises exclusively from saturable erbium absorption rather than thermo-optic or photorefractive effects. We note that the resonance wavelengths drift over a period of several weeks. This may arise from the photorefractive effect in lithium niobate~\cite{langeWidelyNondegenerateNonlinear2025}. However, this has no effect on our results and theoretical model.

\subsubsection{S4. Theoretical Model of Power-Dependent Cavity Dynamics}

To rigorously describe the nonlinear cavity dynamics spanning multiple orders of magnitude of input power (as observed in Fig.~1f of the main text), we develop a steady-state self-consistent model. At the resonant wavelength, the intracavity circulating power ($P_{cav}$) is related to the input power ($P_{in}$) by:
\begin{equation}
P_{cav} = P_{in} \frac{1-r^2}{(1 - r \cdot a(P_{cav}))^2},
\end{equation}
where $r$ is the self-coupling coefficient. The round-trip amplitude transmission $a(P_{cav})$ accounts for both the linear intrinsic loss ($\alpha_{int}$) and the nonlinear saturable absorption of the erbium ions ($\alpha_{Er0}$). To accurately capture the incomplete saturation caused by the non-uniform transverse optical mode profile~\cite{baryaUltraHighQTunable2025a,gritschNarrowOpticalTransitions2022}, we introduce a phenomenological inhomogeneous factor $n$~\cite{dikandeContinuouswavePulseRegimes2017}:
\begin{equation}
a(P_{cav}) = \exp \left[ -\frac{L}{2} \left( \alpha_{int} + \frac{\alpha_{Er0}}{(1 + P_{cav}/P_{sat})^n} \right) \right],
\end{equation}
where $L$ is the cavity perimeter and $P_{sat}$ is the saturation power. The loaded quality factor $Q$ is then given by:
\begin{equation}
Q = \frac{\pi n_g L}{\lambda (1 - r \cdot a)} \sqrt{r \cdot a},
\end{equation}
where $n_g$ is the group index. The theoretical extinction ratio is determined by the off-resonance ($T_{off}$) and on-resonance ($T_{res}$) transmittances. However, in practical devices, the measured extinction ratio is limited by background scattering and polarization crosstalk. Thus, we introduce a background transmission limit ($T_{bg}$) to model the actual extinction ratio:
\begin{equation}
ER_{actual} = 10 \log_{10}\left(\frac{T_{off} + T_{bg}}{T_{res} + T_{bg}}\right).
\end{equation}
To simultaneously fit the drastically different scales of the $Q$ factor (spanning orders of magnitude) and the ER (in decibels), we reconstruct the optimization objective using a variance-weighted least squares approach~\cite{bevington2003data}:
\begin{align}
\mathcal{L} ={}& \sum \left( \frac{\log_{10}Q_{sim} - \log_{10}Q_{exp}}{\sigma_Q} \right)^2 \notag\\
&+ \sum \left( \frac{ER_{sim} - ER_{exp}}{\sigma_{ER}} \right)^2.
\end{align}
Here, $Q_{sim}$ and $ER_{sim}$ represent the simulated values derived from our model, while $Q_{exp}$ and $ER_{exp}$ are the experimentally measured data points. A base-10 logarithmic transformation is applied to the $Q$ factors to linearly scale their variation across multiple orders of magnitude. The terms $\sigma_Q$ and $\sigma_{ER}$ correspond to the standard deviations of the logarithmic experimental $Q$ factors ($\log_{10}Q_{exp}$) and the experimental extinction ratios ($ER_{exp}$), respectively. By scaling the residuals with their corresponding standard deviations, this normalization ensures that both variables—despite possessing entirely different physical units and magnitudes—are weighted equally in the global objective function.

\begin{figure}[htbp]
\centering
\includegraphics[width=\linewidth]{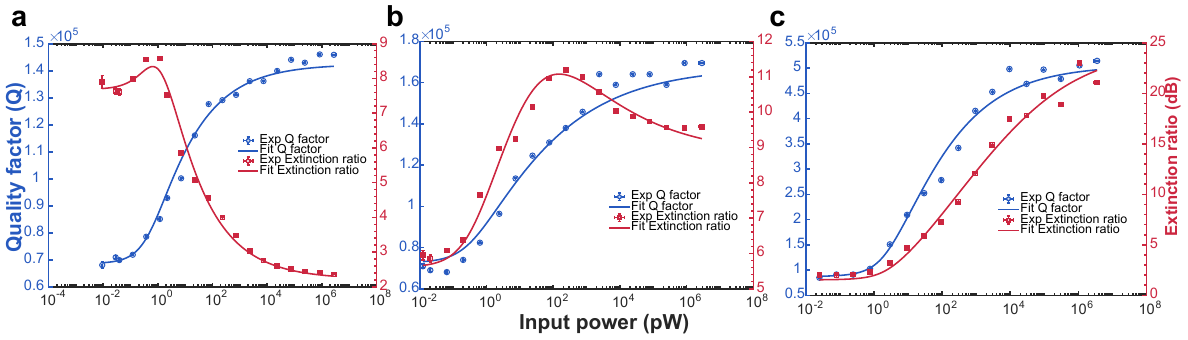}
\caption{\label{figS4} \textbf{Power-dependent cavity dynamics and theoretical fits.} \textbf{a--c}, Simultaneous fits of the loaded quality factor $Q$ (top panels) and extinction ratio ER (bottom panels) as a function of the input power. The data are obtained from three distinct microrings from DEV1 featuring different external coupling rates: $\kappa_{\mathrm{ext}}/2\pi =$ 991.0 MHz \textbf{a}, 684.2 MHz \textbf{b}, and 236.3 MHz \textbf{c}. Solid lines represent the global fits.}
\end{figure}

As shown in Fig.~\ref{figS4}, this model reproduces both $Q$ and ER across the measured power range. The robustness of the model is validated across three distinct microrings on the same 500-ppm chip with varying external coupling rates ($\kappa_{\mathrm{ext}}/2\pi =$ 991.0, 684.2, and 236.3~MHz, respectively). Notably, Fig.~\ref{figS4}a corresponds to the device used in the main text (Fig.~1f). The maximum $Q$ factor of $1.5 \times 10^5$ reached at the highest measured probe power is slightly below the ion-free value of $1.78 \times 10^5$ obtained in Fig.~1d of the main text, because a small residual fraction of the erbium ions remains unsaturated even at the maximum probe power used.

\subsubsection{S5. Detailed Experimental Setup for AFC Preparation and Quantum Storage}

\begin{figure}[htbp]
\centering
\includegraphics[width=0.5\linewidth]{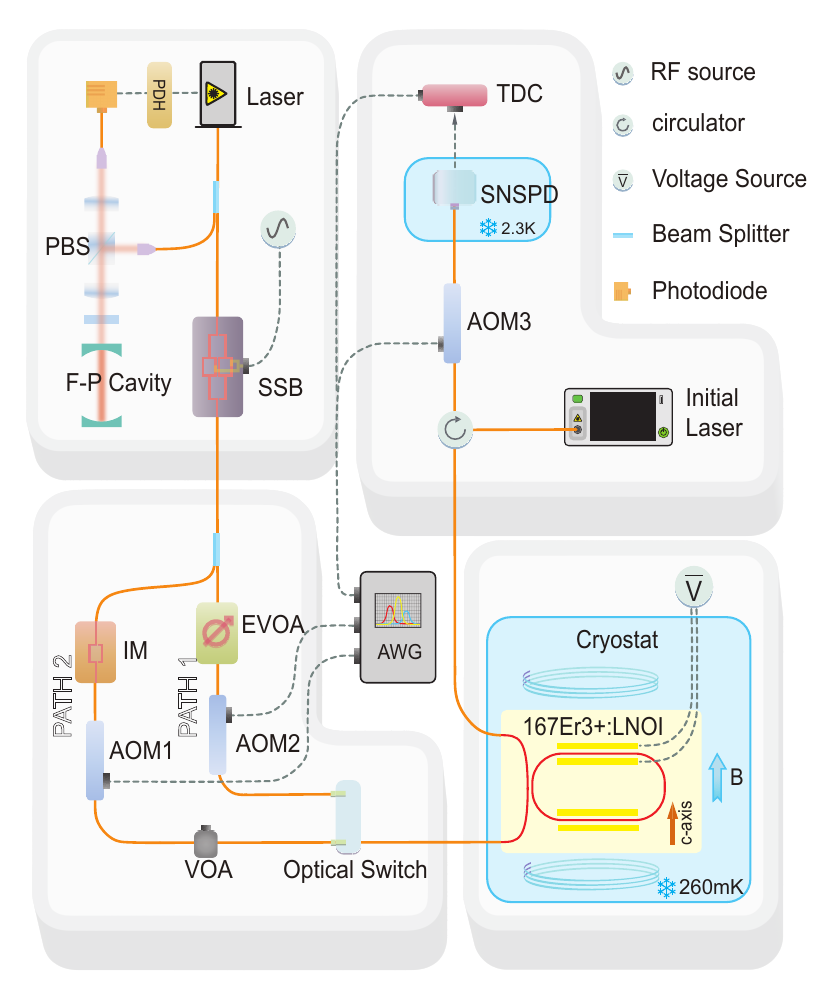}
\caption{\label{figS5} \textbf{Experimental setup for atomic frequency comb (AFC) preparation and quantum storage.} The system comprises laser frequency stabilization by PDH locking to an F-P cavity, frequency shifting with an SSB modulator, optical pulse generation and amplitude control using an IM and AOMs, and electronic timing control using an arbitrary waveform generator (AWG). The $^{167}\text{Er}^{3+}$:TFLN memory device sits in a dilution refrigerator at 260~mK. Echoes are routed via a circulator and gated by AOM3 before being detected by a superconducting nanowire single-photon detector (SNSPD) and recorded by a time-to-digital converter (TDC).}
\end{figure}

To execute the quantum storage protocol, a stable and precisely timed experimental setup is established (Fig.~\ref{figS5}). The primary continuous-wave (CW) laser is first frequency-stabilized using the Pound-Drever-Hall (PDH) technique. A fraction of the laser output is routed to an ultra-stable Fabry--Pérot (F-P) reference cavity featuring a free spectral range (FSR) of 1.5~GHz and a linewidth of less than 40~kHz. By feeding the error signal back to the laser, its frequency is locked to one of the cavity modes, suppressing the laser's linewidth to within 10~kHz. This frequency locking is crucial to prevent laser frequency drift, which would otherwise induce extraneous broadening and degrade the spectral hole-burning resolution.

The main transmitted laser beam passes through a single-sideband (SSB) modulator driven by an RF source. This SSB modulator provides dynamic frequency shifting necessary for sweeping the frequency during AFC preparation and spectral probing. The frequency shift is bypassed for the input photon pulses during the storage phase. Following the SSB, the beam is split into two optical paths using a beam splitter. The first path serves as the strong hole-burning pump. To prevent any leakage light from perturbing the prepared AFC structure, this pump is gated by an acousto-optic modulator (AOM2) to guarantee a sufficiently high extinction ratio, as an optical switch alone cannot provide adequate isolation. On the second path, the light is heavily attenuated using a variable optical attenuator (VOA) to act either as a weak CW probe for mapping the prepared AFC profile, or it is chopped by an intensity modulator (IM) and further gated by AOM1 to generate weak coherent pulses for photon storage.

The hole-burning pump and the signal light are subsequently multiplexed via a fast optical switch, selectively routing them into the $^{167}\text{Er}^{3+}$:TFLN quantum memory according to the experimental sequence. The entire temporal sequence is controlled by an arbitrary waveform generator (AWG), which governs the RF driving signals for the AOMs, the electrical variable optical attenuator, and the optical switch. The AWG simultaneously provides a synchronization reference signal to the time-to-digital converter (TDC).

As discussed in the main text, to reset the ion population prior to the preparation sequence, a separate tunable CW laser (Initial Laser, Santec TSL-570) is employed. This strong initialization beam is injected backward into the memory through a circulator at the output port of the cryostat. During the photon storage and retrieval phase, AOM3 is gated to collect the retrieved optical echoes, which are subsequently detected by a superconducting nanowire single-photon detector (SNSPD) operating at 2.3~K and timestamped by the TDC.

\subsubsection{S6. Optical Coherence of Bulk $^{167}\text{Er}^{3+}$:LN Crystals and TFLN Devices}

\begin{figure}[htbp]
\centering
\includegraphics[width=\linewidth]{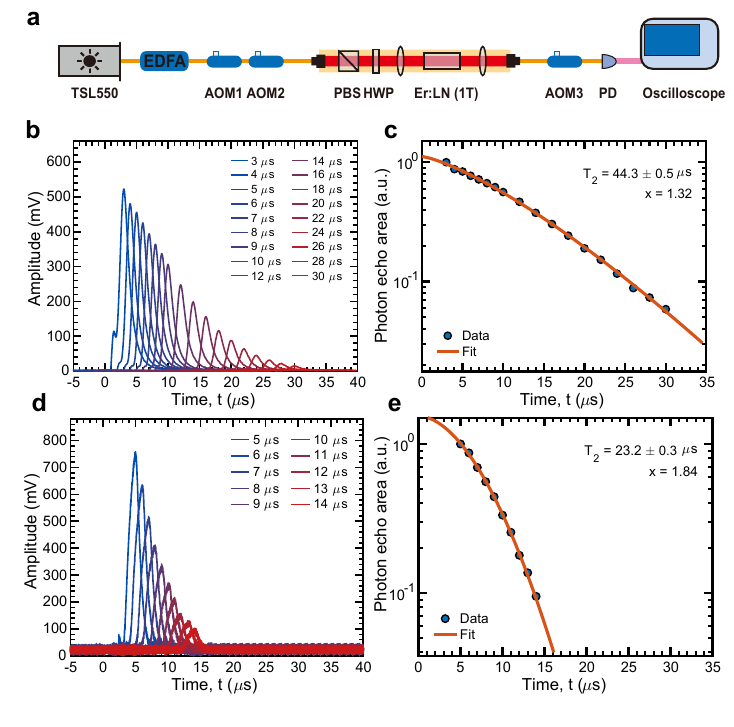}
\caption{\label{figS6} \textbf{Optical coherence of bulk $^{167}\text{Er}^{3+}$:LN crystals.} \textbf{a}, Experimental setup for the Hahn echo measurement at 3.8 K and 1.0 T. \textbf{b,c}, Hahn echo signals and the Mims decay fit for the 50-ppm bulk sample, yielding $T_2 = 44.3(5) \mu\text{s}$. \textbf{d,e}, Corresponding measurements and fit for the 500-ppm bulk sample, yielding $T_2 = 23.2(3) \mu\text{s}$.}
\end{figure}

We characterize the optical coherence time ($T_2$) of the 50-ppm and 500-ppm bulk $^{167}\text{Er}^{3+}$:LN crystals using a standard Hahn echo sequence ($\pi/2 - \tau - \pi - \tau - \text{echo}$). The samples are cooled to 3.8~K with a 1.0~T magnetic field. As depicted in Fig.~\ref{figS6}a, cascaded acousto-optic modulators (AOM1 and AOM2) provide high-extinction pulse chopping, while AOM3 acts as a shutter to protect the photodiode (PD) from saturation during the strong excitation pulses. By tracking the optimized echo area $I(2\tau)$ as a function of the delay time $\tau$ and fitting with a Mims decay model:
\begin{equation}
I(2\tau) = I_0 \exp\left[ -2\left( \frac{2\tau}{T_2} \right)^x \right]
\end{equation}
where $I_0$ is the initial echo area and $x$ is the stretched exponential parameter. From this fit, we extract $T_2 = 44.3(5)~\mu\text{s}$ for the 50-ppm bulk sample (Fig.~\ref{figS6}b,c) and $23.2(3)~\mu\text{s}$ for the 500-ppm bulk sample (Fig.~\ref{figS6}d,e). The reduced coherence at higher doping concentration is attributed to enhanced Er--Er dipolar interactions.

\begin{figure}[htbp]
\centering
\includegraphics[width=\linewidth]{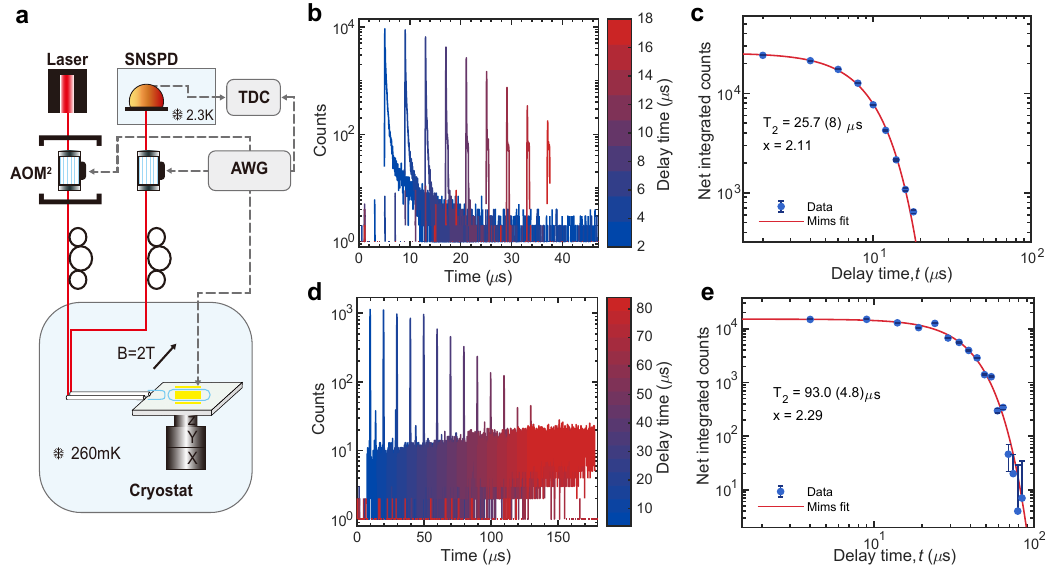}
\caption{\label{figS7} \textbf{Optical coherence of TFLN devices and the effect of thermal annealing.} \textbf{a}, Single-photon-level Hahn echo setup using an SNSPD in a dilution refrigerator (260~mK, 2.0~T). \textbf{b,c}, Hahn echo measurements for the unannealed 50-ppm TFLN device, DEV2 ($T_2 = 25.7(8) \mu\text{s}$). The coherence is limited by etching-induced surface damage. \textbf{d,e}, Hahn echo measurements for the 500-ppm TFLN device, DEV1 with oxygen annealing. The coherence time is extended to $93.0(4.8) \mu\text{s}$.}
\end{figure}

For the integrated TFLN samples, the reduced ensemble volume requires single-photon-level echo detection. The devices are thermalized to 260~mK under a 2.0~T magnetic field. The echoes are detected by an SNSPD, with the signal defined as the integrated counts in the echo window minus the background noise (Fig.~\ref{figS7}a). The unannealed 50-ppm TFLN sample exhibits a $T_2$ of $25.7(8) \mu\text{s}$ (Fig.~\ref{figS7}b,c). Despite the lower temperature and higher magnetic field compared to the bulk measurement, this reduction indicates coherence degradation caused by surface defects, lattice damage, and charge accumulation introduced during the smart-cut and dry-etching processes. Conversely, the 500-ppm TFLN sample---which underwent post-etching oxygen annealing~\cite{baryaUltraHighQTunable2025a}---yields an extended $T_2$ of $93.0(4.8)~\mu$s (Fig.~\ref{figS7}d,e). The beneficial role of annealing is supported by the following comparison. First, the two bulk samples measured under identical conditions (3.8~K, 1.0~T) show that raising the doping from 50 to 500~ppm shortens $T_2$ from $44.3(5)$ to $23.2(3)~\mu$s, as expected from stronger Er--Er dipolar interactions. By this trend alone, the 500-ppm TFLN device should have a shorter $T_2$ than the 50-ppm one. The two TFLN samples, however, show the opposite: measured at the same 260~mK and 2.0~T, the annealed 500-ppm device outperforms the unannealed 50-ppm device. Since the measurement conditions are identical, this reversal is consistent with the beneficial effect of the oxygen anneal applied to DEV1, which mitigates etching-induced lattice damage and oxygen vacancies.

\subsubsection{S7. Spectral Hole Burning and Superhyperfine Interactions}

To characterize the local spin environment of the $^{167}\text{Er}^{3+}$ ions in the TFLN matrix, we perform detailed spectral hole-burning measurements under varying magnetic fields. In this solid-state system, the electron spin of each erbium ion interacts with the surrounding nuclear spins of the host lattice---primarily $^{7}\text{Li}$ and $^{93}\text{Nb}$. Burning a single spectral hole in the inhomogeneously broadened absorption profile therefore also generates additional side-holes at well-defined detunings from the central feature.

\begin{figure}[htbp]
\centering
\includegraphics[width=\linewidth]{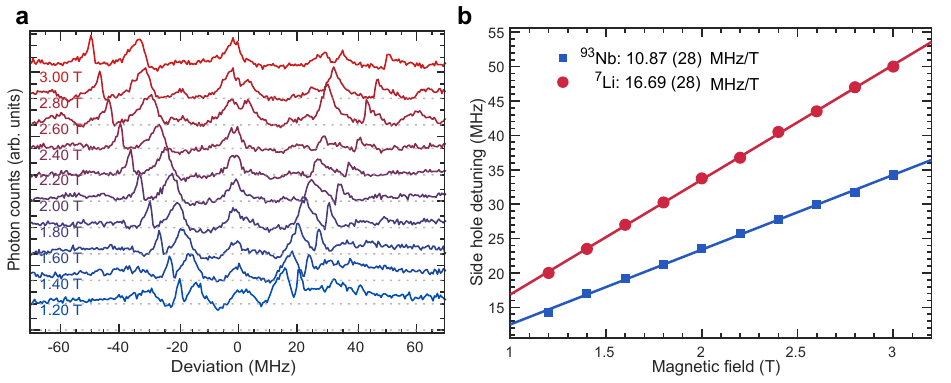}
\caption{\label{figS8} \textbf{Superhyperfine interactions between $^{167}\text{Er}^{3+}$ ions and host Li and Nb nuclei in TFLN.} \textbf{a}, Transmission spectra showing the central spectral hole and associated side holes at various external magnetic fields. \textbf{b}, Dependence of the side-hole detunings on magnetic field.}
\end{figure}

Before describing the measurement, we note that the appearance of a spectral ``hole'' (a localized reduction of ion absorption) as either a peak or a dip in the transmitted spectrum depends on the initial coupling regime of the cavity--ion system. As shown in Fig.~1e,f of the main text, DEV1 operates near critical coupling at low probe powers, with a high extinction ratio and near-zero baseline transmission. When a spectral hole is burned, the local effective absorption decreases, which pushes the cavity back toward an over-coupled state; consequently, the resonant transmission increases and the spectral holes, together with the associated superhyperfine side-holes, naturally appear as peaks in the transmitted photon counts (Fig.~\ref{figS8}a). Conversely, in a cavity with a lower initial extinction ratio that operates in the under-coupled regime, the same reduction in absorption would instead pull the system toward critical coupling and manifest as a transmission dip.

For each measurement a CW pump with an on-chip power of about 82~pW is applied for 10~s to create the central spectral hole, and the resulting spectral feature is then mapped by sweeping a weak probe laser across the resonance. To ensure that the probe does not perturb the prepared hole, its intensity is heavily attenuated to yield approximately 5{,}000 detected photon counts per second, corresponding to a low on-chip probe power of $\sim 6.5$~fW. Fig.~\ref{figS8}a shows the evolution of the hole-burning spectra as the static magnetic field---applied parallel to the crystal's optical $c$-axis---is increased from 1.20~T to 3.00~T. Two distinct sets of side-holes are clearly resolved, symmetrically distributed around the central hole. Extracting their detuning relative to the central frequency and plotting it against the magnetic field (Fig.~\ref{figS8}b) yields a linear dependence, with field-dependent slopes of $10.87(28)$~MHz/T and $16.69(28)$~MHz/T. These values are consistent with the known gyromagnetic ratios of the $^{93}\text{Nb}$ and $^{7}\text{Li}$ nuclei, respectively~\cite{thielOpticalDecoherencePersistent2010,askaraniPersistentAtomicFrequency2020}, confirming that the side-holes originate from the superhyperfine coupling between the $^{167}\text{Er}^{3+}$ electron spin and the TFLN host nuclear spins.

The same side-hole structure directly affects the effective contrast of the prepared AFC. If the side-hole detunings are comparable to the AFC bandwidth but not commensurate with the comb period, the side-holes---which are local transparency windows---tend to fall on the absorption teeth and weaken them. When the magnetic field is tuned so that the side-hole detunings are integer multiples of the AFC tooth spacing $\Delta$, the side-holes instead coincide with the troughs, and the comb contrast is preserved. This is the alignment strategy used in the main text for the targeted storage times.

\subsubsection{S8. Control of AFC Finesse}

\begin{figure}[htbp]
\centering
\includegraphics[width=\linewidth]{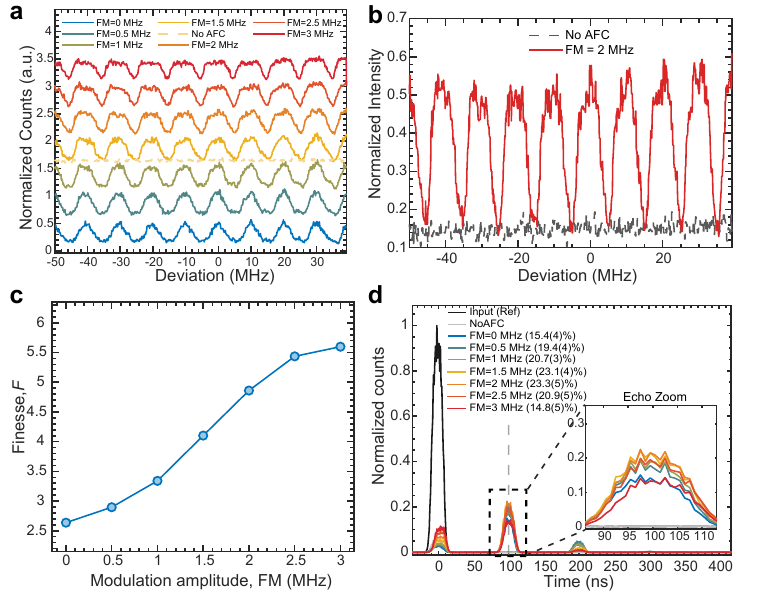}
\caption{\label{figS9} \textbf{Control of AFC finesse via frequency modulation.} \textbf{a}, Transmission spectra of the prepared AFCs at various FM amplitudes ranging from 0 to 3~MHz. The traces are vertically shifted by successive offsets of 0.5 for clarity. \textbf{b}, Zoomed-in transmission spectrum of the high-contrast AFC prepared at an FM amplitude of 2~MHz. \textbf{c}, The extracted comb finesse $F$ as a function of the applied FM amplitude. \textbf{d}, Time-resolved histograms of the retrieved signals at a 100-ns storage time for different FM amplitudes, directly illustrating the dependence of storage efficiency on comb finesse.}
\end{figure}

To verify our theoretical model regarding the dynamic trade-off between comb finesse ($F$) and effective absorption, we characterize the spectral profiles of the AFCs prepared under various frequency modulation (FM) amplitudes. The measurement methodology is identical to that of Fig.~2b in the main text, utilizing DEV1, with a peak on-chip hole-burning power of approximately 260~pW. As shown in Fig.~\ref{figS9}a, increasing the FM amplitude systematically narrows the spectral teeth. For visual clarity, the plotted transmission spectra are vertically shifted by successive increments of 0.5. This demonstrates a near-unity spectral hole-burning efficiency ($\eta_{\text{spectral}} \approx 0.95$). A high-resolution zoomed-in view of the comb structure prepared at FM = 2~MHz is provided in Fig.~\ref{figS9}b.

To quantitatively evaluate the relationship between the applied FM amplitude and the resulting comb finesse (Fig.~\ref{figS9}c), the finesse is calculated as $F = \Delta / \gamma_{\text{tooth}}$, where $\Delta = 10$~MHz is the fixed comb tooth spacing. The effective tooth linewidth ($\gamma_{\text{tooth}}$) is extracted from the measured AFC spectra (Fig.~\ref{figS9}a) by evaluating the width at a specific extinction ratio, which corresponds to the half-maximum of the ion absorption. It is determined from the power-dependent $Q$-factor measurements outlined in Fig.~1f of the main text. Finally, Fig.~\ref{figS9}d displays the corresponding time-resolved histograms of the photon echoes retrieved after a 100-ns storage time for these varying FM configurations. These temporal profiles directly showcase the efficiency evolution governed by the comb finesse, forming the experimental basis for the efficiency optimization curve presented in Fig.~2c of the main text.

\subsubsection{S9. Storage Efficiency at Various Storage Times}

\begin{figure}[htbp]
\centering
\includegraphics[width=0.5\linewidth]{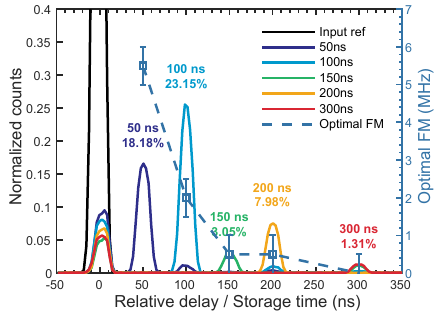}
\caption{\label{figS10} \textbf{Storage efficiency optimized for various predetermined storage times.} Time-resolved histograms of the retrieved photon echoes for storage times ranging from 50~ns to 300~ns. The colored solid lines represent the optimal retrieved signals, with the corresponding optimized storage efficiencies. The dashed dark blue line, referenced to the right axis, indicates the corresponding optimal frequency modulation (FM) amplitude required to maximize the efficiency at each specific storage duration.}
\end{figure}

To map the temporal behavior of the quantum memory, we measure the storage efficiency across a range of predetermined storage times. The emission time of the photon echo, $t_M = 1/\Delta$, is controlled by the AFC tooth spacing $\Delta$; as shown in Fig.~\ref{figS10}, the retrieved echoes appear at the programmed delay for each chosen $\Delta$. For each target storage time we systematically scan the frequency-modulation (FM) amplitude to optimize the comb finesse $F$ and thus the storage efficiency.

The optimized efficiency does not follow a simple exponential decay: distinct local maxima appear at 100~ns and 200~ns. This non-monotonic behavior has the same origin as discussed in Supplementary Section S7: as $\Delta$ is varied, the $^{7}\text{Li}$ and $^{93}\text{Nb}$ superhyperfine side holes alternately align with the transparency troughs and absorption teeth of the comb, periodically minimizing the degradation of the effective comb contrast and thereby reinforcing coherent re-emission at selected storage times.

The dashed dark blue line in Fig.~\ref{figS10} illustrates the inverse relationship between the storage time and the optimal FM amplitude. As the storage time increases, the required tooth spacing narrows, and the FM amplitude must be proportionally reduced in order to keep the theoretical finesse near its optimum and to maintain a sufficiently high spectral hole-burning efficiency ($\eta_{\text{spectral}}$). Beyond 300~ns the optimal efficiency drops sharply. This roll-off is due to the finite spectral hole-burning linewidth of the erbium ensemble. At such narrow tooth spacings ($\sim 2$~MHz) adjacent comb teeth begin to overlap, so that the unpumped background absorption rises and the comb contrast is severely degraded.

\subsubsection{S10. Temporal Multiplexing Capabilities}

\begin{figure}[htbp]
\centering
\includegraphics[width=\linewidth]{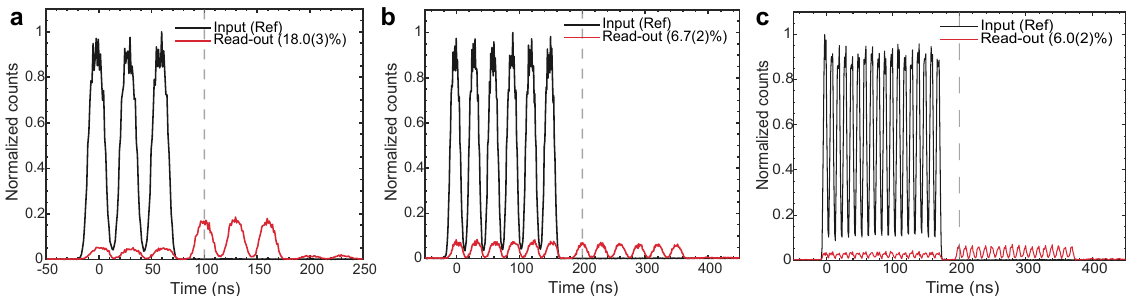}
\caption{\label{figS11} \textbf{Temporally multiplexed quantum storage.} Time-resolved photon counts demonstrating the storage and retrieval of multiple temporal modes. \textbf{a}, Storage of 3 temporal modes using 15-ns pulses with a 100-ns storage time, yielding a collective efficiency of $18.0(3)\%$. \textbf{b}, Storage of 6 modes using 15-ns pulses with a 200-ns storage time, yielding an efficiency of $6.7(2)\%$. \textbf{c}, Storage of 18 modes using compressed 5-ns pulses with an extended 200-ns storage time, achieving an efficiency of $6.0(2)\%$.}
\end{figure}

To systematically evaluate the high-capacity multiplexing capabilities of our memory, we perform temporal-mode multiplexing experiments using various pulse widths, mode counts, and predetermined storage times. The intrinsic broad inhomogeneous bandwidth of the erbium ensemble allows for the storage of short, broadband pulses. As shown in Fig.~\ref{figS11}a, we first utilize 15-ns input pulses within a 100-ns storage window, successfully storing and retrieving 3 distinct temporal modes with a collective efficiency of $18.0(3)\%$. By extending the AFC storage time to 200~ns, we expand the capacity to accommodate 6 temporal modes (using the same 15-ns pulses), resulting in a collective efficiency of $6.7(2)\%$ (Fig.~\ref{figS11}b). To further push the multiplexing limit within this 200-ns window, we compress the input pulse width to 5~ns, which allows us to densely pack and successfully retrieve 18 sequential temporal modes while maintaining a robust collective efficiency of $6.0(2)\%$ (Fig.~\ref{figS11}c).

\subsubsection{S11. Electro-Optic Modulation Simulation}

\begin{figure}[htbp]
\centering
\includegraphics[width=\linewidth]{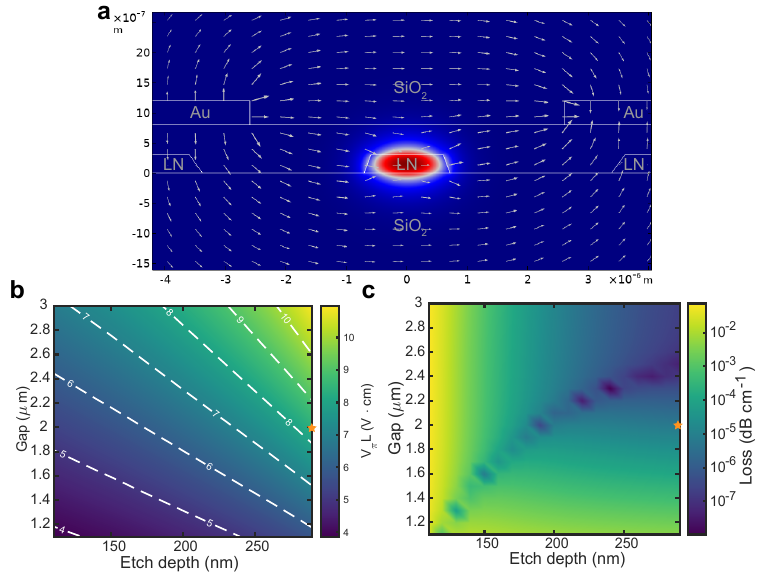}
\caption{\label{figS12} \textbf{Finite-element simulation of the electro-optic microring resonator.} \textbf{a}, Cross-sectional profile of the fundamental TE optical mode overlaid with the simulated electrostatic field distribution (white arrows). \textbf{b,c}, Simulated half-wave voltage-length product ($V_\pi L$) and metal-induced propagation loss as functions of the electrode gap and etching depth. The yellow stars indicate the design parameters used in this work.}
\end{figure}

To optimize the electro-optic tuning of the microring resonator while minimizing metal-induced absorption, we perform finite-element simulations using \textsc{COMSOL Multiphysics}. Fig.~\ref{figS12}a displays the simulated fundamental transverse-electric (TE) optical mode tightly confined within the ridge waveguide, overlaid with the electrostatic field distribution (white arrows) generated by the micro-electrodes. We systematically mapped the half-wave voltage-length product ($V_\pi L$, Fig.~\ref{figS12}b) and the optical propagation loss (Fig.~\ref{figS12}c) across various electrode gaps and waveguide etching depths; the yellow stars mark the parameters of our fabricated device. For this optimized geometric design, the simulated EO-induced index change is derived from the Pockels relation
\begin{equation}
\Delta n = \frac{E_{\mathrm{avg}} n_e^3 r_{33}}{2},
\end{equation}
where $E_{\mathrm{avg}}$ is the optical-mode-weighted electric field generated by the electrodes. The simulated index tuning efficiency is $\Delta n/\Delta V = 8.79\times10^{-6}~\mathrm{V}^{-1}$. The half-wave voltage-length product is then calculated from the condition $\Delta n L=\lambda/2$ as
\begin{equation}
V_\pi L = \frac{\lambda}{2(\Delta n/\Delta V)}.
\end{equation}
Here we use the $\lambda/2$ phase-shift condition because the microring tuning corresponds to a single-ended phase modulation. This differs from the commonly used push-pull MZI definition, where the effective differential phase shift is doubled and the corresponding expression contains $\lambda/4$. Using the operating wavelength, this gives $V_\pi L = 8.71~\mathrm{V}\cdot\mathrm{cm}$. The theoretical resonance frequency tuning efficiency ($\Delta \nu / \Delta V$) can be derived as:
\begin{equation}
\frac{\Delta \nu}{\Delta V} = \frac{\mathrm{FSR}\, L_e}{2 V_\pi L}.
\end{equation}
Given the cavity free spectral range ($\text{FSR} = 50~GHz$) and the effective electrode length ($L_e = 2254.7\ \mu\text{m}$), the theoretical tuning efficiency is calculated to be 647 MHz/V. This simulated value is consistent with the experimental tuning performance demonstrated in the main text (Fig.~3). Furthermore, the simulation confirms that the electrode-induced optical absorption loss at the chosen design point is only $6.6 \times 10^{-4}$ dB/cm. This value is nearly three orders of magnitude smaller than the intrinsic waveguide propagation loss (0.325 dB/cm), ensuring that the integration of the tuning electrodes introduces negligible excess loss to the stored single photons.

\subsubsection{S12. Electro-Optic Response Characteristics}

\begin{figure}[htbp]
\centering
\includegraphics[width=0.8\linewidth]{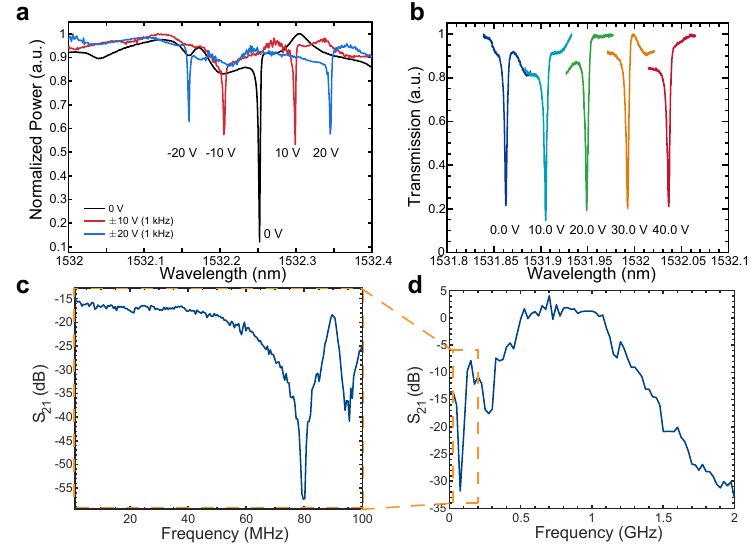}
\caption{\label{figS13} \textbf{Electro-optic tuning and modulation frequency response.} \textbf{a,b}, Electro-optic tuning of the microring resonance at room temperature (RT) and 4~K, respectively. \textbf{c,d}, Electro-optic frequency response ($S_{21}$) measured by a vector network analyzer at RT and 4~K, respectively.}
\end{figure}

We characterize the EO tuning efficiency of the TFLN microring at both room temperature (RT) and 4~K. At RT, lithium niobate devices typically exhibit prominent DC drift due to the photorefractive effect and charge migration; to circumvent this, we apply bipolar square-wave voltages ($\pm 10$~V and $\pm 20$~V) and sweep the transmission spectra (Fig.~\ref{figS13}a), yielding a tuning efficiency of 4.65~pm/V (594~MHz/V). At 4~K the DC drift is strongly suppressed~\cite{warnerDCstableThinfilmLithium2025}, allowing direct application of static DC voltages from 0~V to 40~V (Fig.~\ref{figS13}b). The extracted cryogenic tuning efficiency is 4.34~pm/V (554~MHz/V), consistent with the theoretical value of 647~MHz/V from our finite-element simulations.

We further evaluate the high-speed dynamic performance by measuring the EO frequency response ($S_{21}$) with a vector network analyzer (Fig.~\ref{figS13}c,d). The device exhibits a baseband 3-dB bandwidth exceeding 60~MHz. A sharp dip near 80~MHz is attributed to electrical impedance mismatch and microwave cable reflections in the cryogenic setup. Beyond the baseband, a prominent resonance-enhanced modulation band centered near 700~MHz provides a 6-dB passband bandwidth greater than 400~MHz. These broadband EO characteristics establish the hardware foundation for the frequency-selective storage and routing demonstrated in the main text.

\subsubsection{S13. Voltage-Dependent Suppression of Residual Echoes in DEV1}

\begin{figure}[htbp]
\centering
\includegraphics[width=0.5\linewidth]{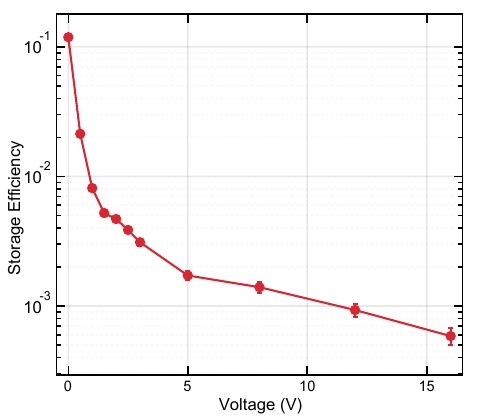}
\caption{\label{figS14} \textbf{Voltage-dependent suppression of residual echoes in DEV1.} An additional bias voltage is applied during storage and readout after AFC preparation, detuning the microring resonance from the stored-signal frequency. The plotted quantity is the residual echo efficiency $\eta_{\mathrm{res}}=N_{\mathrm{echo}}/N_{\mathrm{in}}$ under the detuned readout condition, which decreases rapidly with increasing voltage offset. The normalized inter-channel crosstalk is obtained by dividing $\eta_{\mathrm{res}}$ by the matched storage efficiency at zero bias.}
\end{figure}

We characterize the voltage-dependent suppression of residual echoes in DEV1 by applying an additional voltage offset to the chip after the AFC preparation step. During AFC preparation, both the AFC pump field and the weak coherent signal are set to the microring resonance at zero applied voltage. After the AFC is prepared, an additional bias voltage $V_{\mathrm{bias}}$ is applied during the storage and readout sequence. For $V_{\mathrm{bias}}=0$, the readout cavity configuration remains matched to the signal frequency, corresponding to the normal retrieval condition. For nonzero $V_{\mathrm{bias}}$, the microring resonance is shifted away from the stored-signal frequency, so that the retrieved echo is no longer resonantly enhanced at the signal frequency. The residual echo efficiency under the detuned readout condition is defined as:
\begin{equation}
\eta_{\mathrm{res}}=
\frac{N_{\mathrm{echo}}}{N_{\mathrm{in}}},
\end{equation}
where $N_{\mathrm{echo}}$ is the photon count integrated over the expected echo time window and $N_{\mathrm{in}}$ is the incident photon number of the input signal pulse. The corresponding normalized inter-channel crosstalk is defined as
\begin{equation}
\chi=
\frac{\eta_{\mathrm{res}}}{\eta_{0}},
\end{equation}
where $\eta_{0}$ is the matched storage efficiency at zero bias. As shown in Fig.~\ref{figS14}, the retrieval efficiency decreases rapidly with increasing voltage offset, from $\sim 10^{-1}$ under the matched condition to a few $10^{-3}$ at $V_{\mathrm{bias}}\approx 3~\mathrm{V}$, indicating that even a moderate cavity detuning substantially suppresses the storage and echo process. Increasing the voltage offset to $16~\mathrm{V}$ further suppresses the residual echo efficiency to below $10^{-3}$, corresponding to a normalized crosstalk below $10^{-2}$.

\subsubsection{S14. Frequency-Selective Storage and Crosstalk Characterization}

\begin{figure}[htbp]
\centering
\includegraphics[width=\linewidth]{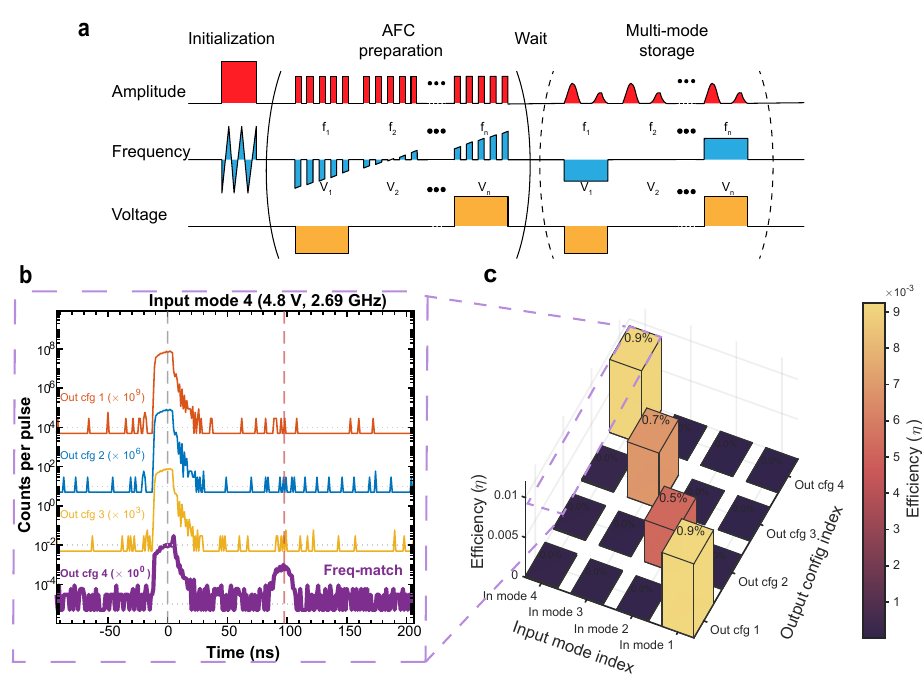}
\caption{\label{figS15} \textbf{Frequency-selective storage and low-crosstalk demonstration in DEV2.} \textbf{a}, Experimental sequence for frequency-selective storage. The cavity is electro-optically tuned to predetermined frequencies ($f_n$) via distinct DC voltages ($V_n$) to sequentially prepare independent AFCs. \textbf{b}, Time-resolved histograms illustrating the crosstalk measurement for Input Mode 4 (4.8 V, 2.69 GHz). A pronounced echo is retrieved only under the frequency-matched output configuration, whereas mismatched configurations yield signals at the noise floor. \textbf{c}, 3D bar chart summarizing the storage efficiencies and inter-channel crosstalk across a four-channel configuration. Crosstalk for mismatched channels is suppressed below $10^{-2}$.}
\end{figure}

The experimental sequence for frequency-selective storage is outlined in Fig.~\ref{figS15}a. Following the population initialization, we apply distinct DC voltages to the microring electrodes. This shifts the resonance to target frequency channels, where independent AFCs are sequentially prepared using matching SSB frequency shifts. During the subsequent storage phase, the readout voltage is set to one of the corresponding configurations.

We first employ the 50-ppm doped device (DEV2), which combines a weaker ion-cavity coupling with a deliberately reduced external coupling rate, yielding a higher loaded quality factor of $3.6 \times 10^5$ and a correspondingly narrower cavity linewidth. To quantify the channel crosstalk, we measure the echoed signals under all permutations of input modes and output configurations. As depicted in Fig.~\ref{figS15}b, taking Input Mode 4 as an example, when the input photon's frequency precisely matches the electro-optically tuned cavity resonance and its corresponding AFC (Output config 4), the echoed signal is successfully retrieved. Conversely, if the input mode and output configuration are mismatched, the cavity resonance is shifted away, and the echo signals are suppressed to the noise floor. The crosstalk from mismatched frequency modes is heavily suppressed to below $10^{-2}$, limited by the detector noise floor.

\begin{figure}[htbp]
\centering
\includegraphics[width=\linewidth]{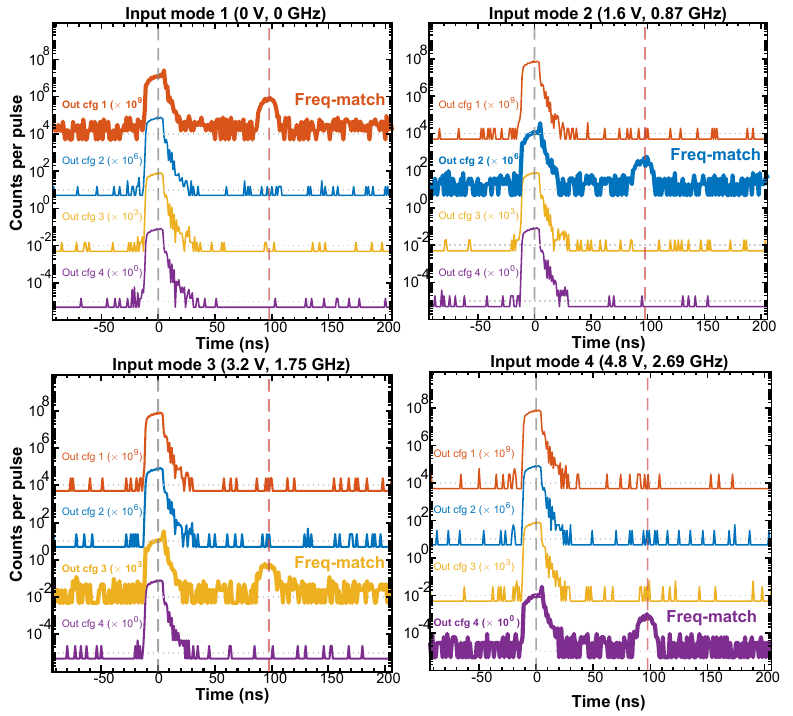}
\caption{\label{figS16} \textbf{Comprehensive crosstalk characterization for DEV2.} Time-resolved photon counts for all four input frequency channels (0~GHz, 0.87~GHz, 1.75~GHz, 2.69~GHz) under four different output voltage configurations. The frequency-matched combinations yield distinct retrieval echoes, while all mismatched combinations remain at the noise floor.}
\end{figure}

As summarized in the 3D bar chart (Fig.~\ref{figS15}c), we demonstrated four-channel frequency-selective storage by applying static voltages of 0~V, 1.6~V, 3.2~V, and 4.8~V, which correspond to relative frequency shifts of 0~GHz, 0.87~GHz, 1.75~GHz, and 2.69~GHz, respectively. The resulting storage efficiencies for these four independent channels are 0.9\%, 0.7\%, 0.5\%, and 0.9\%, maintaining consistent performance with negligible excess loss induced by the EO tuning. The time-resolved measurements corresponding to all 16 permutations of the 4 input modes and 4 output configurations in DEV2 are fully shown in Fig.~\ref{figS16}.

\subsubsection{S15. Quantum Storage Efficiency of DEV2}

\begin{figure}[htbp]
\centering
\includegraphics[width=\linewidth]{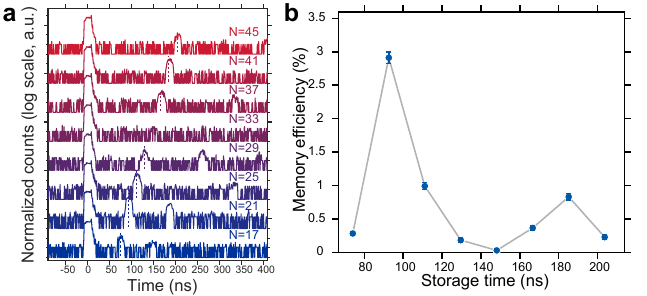}
\caption{\label{figS17} \textbf{Storage efficiency of DEV2 at various storage times.} \textbf{a}, Stacked time-resolved histograms demonstrating the storage and retrieval of weak coherent pulses (average photon number $\mu = 0.768(4)$ per pulse) at various predetermined storage times. $N$ denotes the number of AFC teeth; the total comb bandwidth is fixed at 216~MHz, and the tooth spacing is varied by changing $N$. The vertical dashed lines indicate the designed echo retrieval time for each configuration. Higher order echo signals are visible as well. \textbf{b}, Measured storage efficiency as a function of the storage time.}
\end{figure}

In addition to the efficiency optimization for DEV1 with 500~ppm doping presented in the main text, we also systematically investigate the storage performance of DEV2 across various predetermined storage times. These measurements are conducted under a static magnetic field of 2.0~T. The total AFC bandwidth is fixed at 216~MHz, and the storage time ($T = 1/\Delta f$) is tuned by varying the number of comb teeth $N$, which changes the tooth spacing $\Delta f$ accordingly; the designed echo retrieval times are marked by the vertical dashed lines in Fig.~\ref{figS17}a. Fig.~\ref{figS17}a displays the stacked time-resolved histograms of the retrieved photon echoes. The stored photons are retrieved at the exact programmed temporal delays. Fig.~\ref{figS17}b plots the corresponding storage efficiency as a function of the programmed storage time. Consistent with the physical mechanism described previously, the efficiency exhibits a strongly non-monotonic trend due to the complex local spin environment. Specifically, we observe pronounced local efficiency maxima at $T = 92.6$~ns and $T = 185.2$~ns, reaching storage efficiencies of $2.91(9)\%$ and $0.83(5)\%$, respectively. This enhancement is attributed to the alignment of the superhyperfine side-hole structure with the AFC spectrum. At a magnetic field of 2.0~T, the Zeeman splittings of both the $^{7}\text{Li}$ and $^{93}\text{Nb}$ nuclei create spectral side-holes whose detunings simultaneously coincide with the harmonics of the AFC tooth spacing for these two specific storage times. This alignment places the superhyperfine side holes near the transparency troughs of the AFC and prevents them from overlapping the residual absorption teeth, thereby preserving the effective comb contrast and enhancing the collective re-emission.

\subsubsection{S16. Dynamic Frequency Routing of DEV2}

\begin{figure}[htbp]
\centering
\includegraphics[width=0.8\linewidth]{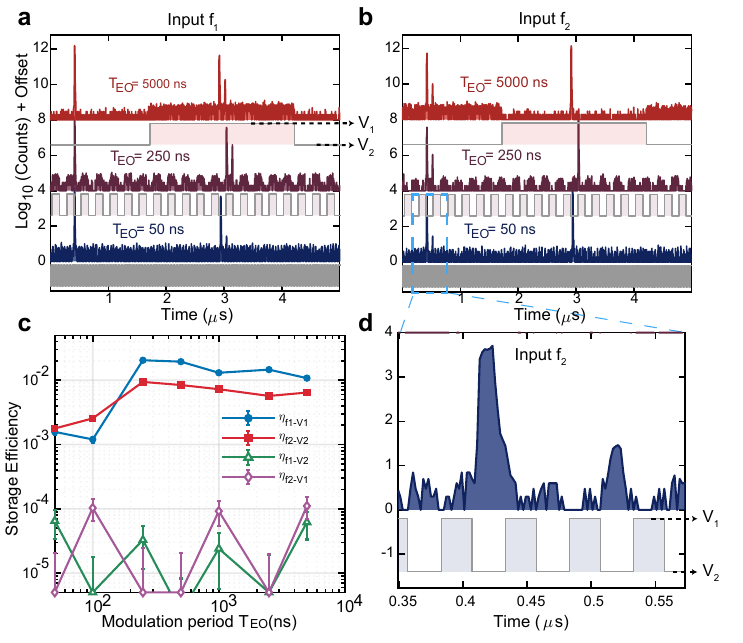}
\caption{\label{figS18} \textbf{Dynamic frequency routing of DEV2.} \textbf{a,b}, Time-resolved retrieval histograms for input frequencies $f_1$ and $f_2$. Echoes are retrieved when the input frequency matches the readout voltage configuration. \textbf{c}, Matched storage efficiencies ($\eta_{f_1-V_1}$, $\eta_{f_2-V_2}$) and unwanted retrieval efficiency ($\eta_{f_1-V_2}$, $\eta_{f_2-V_1}$) as a function of the electro-optic modulation period $T_{\mathrm{EO}}$. \textbf{d}, Enlarged view of the $f_2$ trace at $T_{\mathrm{EO}}=50$~ns, corresponding to the dashed region in \textbf{b}.}
\end{figure}

We further characterize dynamic frequency-selective storage in DEV2, which has a lower erbium concentration and a higher loaded quality factor than DEV1. The experimental sequence follows the dynamic routing measurement described in the main text. The magnetic field is set to 1.95~T and the AFC storage time is 95~ns. A square-wave voltage modulation with amplitude $\pm 0.8$~V is applied to the microring electrodes to switch the cavity resonance between two frequency configurations. The two input frequencies, denoted as $f_1$ and $f_2$, are generated using SSB modulation frequencies of 600 and 1390~MHz relative to the AFC preparation laser, respectively. The input pulses have a duration of 10~ns. Fig.~\ref{figS18}a and b show the time-resolved retrieval histograms when the input is set to $f_1$ and $f_2$, respectively. Similar to the behavior observed in the main-text experiment, the two input pulses are timed such that their echoes fall into different readout voltage configurations. An echo is predominantly retrieved when the input frequency matches the cavity configuration at the readout time. For the $f_1$ input, the matched retrieval under the $V_1$ configuration gives the storage efficiency $\eta_{f_1\text{-}V_1}$, while the signal measured under the mismatched $V_2$ configuration is denoted as the unwanted retrieval efficiency $\eta_{f_1\text{-}V_2}$. Conversely, for the $f_2$ input, $\eta_{f_2\text{-}V_2}$ is the matched storage efficiency, whereas $\eta_{f_2\text{-}V_1}$ quantifies unwanted retrieval under the mismatched $V_1$ configuration.

Fig.~\ref{figS18}c summarizes the matched and unwanted retrieval efficiencies defined above for different modulation periods. The normalized inter-channel crosstalk is defined as $\chi_{f_1\text{-}V_2}=\eta_{f_1\text{-}V_2}/\eta_{f_1\text{-}V_1}$ and $\chi_{f_2\text{-}V_1}=\eta_{f_2\text{-}V_1}/\eta_{f_2\text{-}V_2}$. Owing to the lower doping concentration and the narrower linewidth associated with the higher-$Q_{\mathrm{loaded}}$ microring, the unwanted retrieval efficiencies remain below $10^{-4}$, corresponding to inter-channel crosstalk at the $10^{-2}$ level, limited by noise counts. Fig.~\ref{figS18}d shows a zoomed-in view of the $T_{\mathrm{EO}}=50$~ns trace for the $f_2$ input, illustrating frequency-selective retrieval under the fastest modulation condition tested here.

\subsubsection{S17. The Integrated Entangled Photon-pair Source}

\begin{figure}[htbp]
\centering
\includegraphics[width=0.5\linewidth]{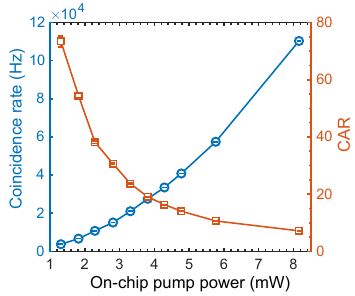}
\caption{\label{figS19} \textbf{Performance of the integrated silicon nitride (SiN) entanglement source.} Measured coincidence rate of the generated signal-idler photon pairs (blue circles, left axis) and the corresponding coincidence-to-accidental ratio (CAR, orange squares, right axis) as a function of the on-chip pump power.}
\end{figure}

We characterize the brightness and correlation quality of the time-energy entangled photons generated via spontaneous four-wave mixing (SFWM) in our integrated silicon nitride (SiN) microring resonator by measuring the photon-pair generation metrics across varying pump powers. As depicted in Fig.~\ref{figS19}, increasing the on-chip pump power leads to a quadratic increase in the coincidence rate, reaching a maximum measured value of approximately $1.1 \times 10^5$~Hz, reflecting the fundamental scaling of the SFWM process, while the coincidence-to-accidental ratio (CAR) drops as multi-pair emission and uncorrelated background noise become more prominent. For the entanglement storage experiments presented in the main text, we select an on-chip pump power of 3.8~mW that provides a sufficiently high coincidence rate of $2.8 \times 10^4$~Hz while maintaining a CAR of about 19. Combined with the TFLN platform's compatibility with integrated electro-optic modulators, on-chip photon-pair sources, and SNSPDs~\cite{zhaoHighQualityEntangled2020,sayemLithiumniobateoninsulatorWaveguideintegratedSuperconducting2020,lomonteSinglephotonDetectionCryogenic2021,colangeloMolybdenumSilicideSuperconducting2024}, the quantum memory demonstrated here provides a practical route toward fully integrated quantum repeater nodes.

\subsubsection{S18. Active Feedback on Quantum Memory for Long-Term Stability}

The entanglement storage experiment requires extended data acquisition for photon-pair coincidence measurements, making long-term stability of the memory essential. Optimal cavity-enhanced storage requires maintaining spectral alignment between the cavity resonance and the signal frequency. In practice, however, lithium niobate devices exhibit a slow resonance drift over time, attributed to the photorefractive effect~\cite{langeWidelyNondegenerateNonlinear2025} and residual DC drift~\cite{warnerDCstableThinfilmLithium2025} caused by slow charge migration and relaxation processes.

\begin{figure}[htbp]
\centering
\includegraphics[width=0.6\linewidth]{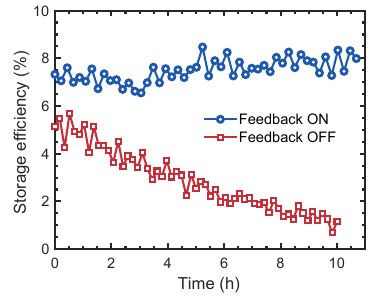}
\caption{\label{figS20} \textbf{Long-term stability of the quantum memory with active voltage feedback.} The storage efficiency is monitored over a 10-hour period. Blue circles represent the stabilized efficiency with the active voltage feedback engaged; red squares show the continuous degradation when the feedback is turned off. Each data point is accumulated over a 10-minute interval.}
\end{figure}

To counteract this drift, we implemented an active voltage feedback loop that periodically monitors the transmitted single counts of the signal photons and dynamically adjusts the DC bias voltage applied to the microring electrodes, re-centering the cavity resonance to the signal frequency. As demonstrated in Fig.~\ref{figS20}, we tracked the storage efficiency over a 10-hour window. When the feedback is disabled (red squares), the cavity gradually drifts out of resonance and the storage efficiency degrades from approximately 5.5\% to near 1\%. With the feedback engaged (blue circles), the cavity resonance remains aligned with the signal frequency, maintaining a stable storage efficiency of $7.5(5)\%$ throughout the entire 10-hour duration.

\bibliographystyle{apsrev4-2}
\bibliography{LNQMbib}

@article{afzeliusImpedancematchedCavityQuantum2010,
  title = {Impedance-Matched Cavity Quantum Memory},
  author = {Afzelius, Mikael and Simon, Christoph},
  year = 2010,
  month = aug,
  journal = {Physical Review A},
  volume = {82},
  number = {2},
  pages = {022310},
  issn = {1050-2947, 1094-1622},
  doi = {10.1103/PhysRevA.82.022310},
  urldate = {2024-07-27},
  copyright = {http://link.aps.org/licenses/aps-default-license},
  langid = {english}
}

@article{afzeliusMultimodeQuantumMemory2009,
  title = {Multimode Quantum Memory Based on Atomic Frequency Combs},
  author = {Afzelius, Mikael and Simon, Christoph and {de Riedmatten}, Hugues and Gisin, Nicolas},
  year = 2009,
  month = may,
  journal = {Physical Review A},
  volume = {79},
  number = {5},
  eprint = {0805.4164},
  primaryclass = {quant-ph},
  pages = {052329},
  issn = {1050-2947, 1094-1622},
  doi = {10.1103/PhysRevA.79.052329},
  urldate = {2024-03-30},
  archiveprefix = {arXiv},
  langid = {english}
}

@article{askaraniPersistentAtomicFrequency2020,
  title = {Persistent Atomic Frequency Comb Based on {{Zeeman}} Sub-Levels of an Erbium-Doped Crystal Waveguide},
  author = {Askarani, Mohsen Falamarzi and Lutz, Thomas and Puigibert, Marcelli Grimau and Sinclair, Neil and Oblak, Daniel and Tittel, Wolfgang},
  year = 2020,
  month = feb,
  journal = {Journal of the Optical Society of America B},
  volume = {37},
  number = {2},
  eprint = {1907.07780},
  primaryclass = {quant-ph},
  pages = {352},
  issn = {0740-3224, 1520-8540},
  doi = {10.1364/JOSAB.373100},
  urldate = {2024-11-19},
  archiveprefix = {arXiv},
  langid = {english}
}

@article{askaraniStorageReemissionHeralded2019,
  title = {Storage and {{Reemission}} of {{Heralded Telecommunication-Wavelength Photons Using}} a {{Crystal Waveguide}}},
  author = {Askarani, Mohsen Falamarzi and Puigibert, Marcel.li Grimau and Lutz, Thomas and Verma, Varun B. and Shaw, Matthew D. and Nam, Sae Woo and Sinclair, Neil and Oblak, Daniel and Tittel, Wolfgang},
  year = 2019,
  month = may,
  journal = {Physical Review Applied},
  volume = {11},
  number = {5},
  pages = {054056},
  issn = {2331-7019},
  doi = {10.1103/PhysRevApplied.11.054056},
  urldate = {2024-11-05},
  langid = {english}
}

@article{assumpcaoThinFilmLithium2024,
  title = {A Thin Film Lithium Niobate Near-Infrared Platform for Multiplexing Quantum Nodes},
  author = {Assumpcao, Daniel and Renaud, Dylan and Baradari, Aida and Zeng, Beibei and {De-Eknamkul}, Chawina and Xin, C. J. and {Shams-Ansari}, Amirhassan and Barton, David and Machielse, Bartholomeus and Loncar, Marko},
  year = 2024,
  month = dec,
  journal = {Nature Communications},
  volume = {15},
  number = {1},
  pages = {10459},
  issn = {2041-1723},
  doi = {10.1038/s41467-024-54541-2},
  urldate = {2025-05-06},
  langid = {english}
}

@article{craiciuMultifunctionalOnchipStorage2021,
  title = {Multifunctional On-Chip Storage at Telecommunication Wavelength for Quantum Networks},
  author = {Craiciu, Ioana and Lei, Mi and Rochman, Jake and Bartholomew, John G. and Faraon, Andrei},
  year = 2021,
  month = jan,
  journal = {Optica},
  volume = {8},
  number = {1},
  pages = {114--121},
  publisher = {Optica Publishing Group},
  issn = {2334-2536},
  doi = {10.1364/OPTICA.412211},
  urldate = {2024-08-23},
  copyright = {\copyright{} 2021 Optical Society of America},
  langid = {english}
}

@article{craiciuNanophotonicQuantumStorage2019,
  title = {Nanophotonic {{Quantum Storage}} at {{Telecommunication Wavelength}}},
  author = {Craiciu, Ioana and Lei, Mi and Rochman, Jake and Kindem, Jonathan M. and Bartholomew, John G. and Miyazono, Evan and Zhong, Tian and Sinclair, Neil and Faraon, Andrei},
  year = 2019,
  month = aug,
  journal = {Physical Review Applied},
  volume = {12},
  number = {2},
  pages = {024062},
  issn = {2331-7019},
  doi = {10.1103/PhysRevApplied.12.024062},
  urldate = {2024-07-21},
  langid = {english}
}

@article{davidsonImprovedLightmatterInteraction2020,
  title = {Improved Light-Matter Interaction for Storage of Quantum States of Light in a Thulium-Doped Crystal Cavity},
  author = {Davidson, Jacob H. and Lefebvre, Pascal and Zhang, Jun and Oblak, Daniel and Tittel, Wolfgang},
  year = 2020,
  month = apr,
  journal = {Physical Review A},
  volume = {101},
  number = {4},
  pages = {042333},
  issn = {2469-9926, 2469-9934},
  doi = {10.1103/PhysRevA.101.042333},
  urldate = {2025-05-30},
  langid = {english}
}

@article{dibosAtomicSourceSingle2018,
  title = {Atomic {{Source}} of {{Single Photons}} in the {{Telecom Band}}},
  author = {Dibos, A. M. and Raha, M. and Phenicie, C. M. and Thompson, J. D.},
  year = 2018,
  month = jun,
  journal = {Physical Review Letters},
  volume = {120},
  number = {24},
  pages = {243601},
  issn = {0031-9007, 1079-7114},
  doi = {10.1103/PhysRevLett.120.243601},
  urldate = {2024-10-28},
  langid = {english}
}

@article{duanLongdistanceQuantumCommunication2001,
  title = {Long-Distance Quantum Communication with Atomic Ensembles and Linear Optics},
  author = {Duan, L.-M. and Lukin, M. D. and Cirac, J. I. and Zoller, P.},
  year = 2001,
  month = nov,
  journal = {Nature},
  volume = {414},
  number = {6862},
  pages = {413--418},
  publisher = {Nature Publishing Group},
  issn = {1476-4687},
  doi = {10.1038/35106500},
  urldate = {2024-09-13},
  copyright = {2001 Macmillan Magazines Ltd.},
  langid = {english}
}

@article{duttaIntegratedPhotonicPlatform2020,
  title = {Integrated {{Photonic Platform}} for {{Rare-Earth Ions}} in {{Thin Film Lithium Niobate}}},
  author = {Dutta, Subhojit and Goldschmidt, Elizabeth A. and Barik, Sabyasachi and Saha, Uday and Waks, Edo},
  year = 2020,
  month = jan,
  journal = {Nano Letters},
  volume = {20},
  number = {1},
  pages = {741--747},
  issn = {1530-6984, 1530-6992},
  doi = {10.1021/acs.nanolett.9b04679},
  urldate = {2024-10-08},
  copyright = {https://doi.org/10.15223/policy-029},
  langid = {english}
}

@article{gritschNarrowOpticalTransitions2022,
  title = {Narrow {{Optical Transitions}} in {{Erbium-Implanted Silicon Waveguides}}},
  author = {Gritsch, Andreas and Weiss, Lorenz and Fr{\"u}h, Johannes and Rinner, Stephan and Reiserer, Andreas},
  year = 2022,
  month = oct,
  journal = {Physical Review X},
  volume = {12},
  number = {4},
  pages = {041009},
  issn = {2160-3308},
  doi = {10.1103/PhysRevX.12.041009},
  urldate = {2024-07-25},
  langid = {english}
}

@article{jiangQuantumStorageEntangled2023,
  title = {Quantum Storage of Entangled Photons at Telecom Wavelengths in a Crystal},
  author = {Jiang, Ming-Hao and Xue, Wenyi and He, Qian and An, Yu-Yang and Zheng, Xiaodong and Xu, Wen-Jie and Xie, Yu-Bo and Lu, Yanqing and Zhu, Shining and Ma, Xiao-Song},
  year = 2023,
  month = nov,
  journal = {Nature Communications},
  volume = {14},
  number = {1},
  pages = {6995},
  publisher = {Nature Publishing Group},
  issn = {2041-1723},
  doi = {10.1038/s41467-023-42741-1},
  urldate = {2024-04-11},
  copyright = {2023 The Author(s)},
  langid = {english}
}

@article{jiangRareEarthimplantedLithium2019,
  title = {Rare Earth-Implanted Lithium Niobate: {{Properties}} and on-Chip Integration},
  shorttitle = {Rare Earth-Implanted Lithium Niobate},
  author = {Jiang, Xiaodong and Pak, Dongmin and Nandi, Arindam and Xuan, Yi and Hosseini, Mahdi},
  year = 2019,
  month = aug,
  journal = {Applied Physics Letters},
  volume = {115},
  number = {7},
  pages = {071104},
  issn = {0003-6951, 1077-3118},
  doi = {10.1063/1.5098316},
  urldate = {2025-06-30},
  langid = {english}
}

@article{jobezCavityenhancedStorageOptical2014,
  title = {Cavity-Enhanced Storage in an Optical Spin-Wave Memory},
  author = {Jobez, P and Usmani, I and Timoney, N and Laplane, C and Gisin, N and Afzelius, M},
  year = 2014,
  month = aug,
  journal = {New Journal of Physics},
  volume = {16},
  number = {8},
  pages = {083005},
  issn = {1367-2630},
  doi = {10.1088/1367-2630/16/8/083005},
  urldate = {2025-06-01},
  copyright = {http://iopscience.iop.org/info/page/text-and-data-mining},
  langid = {english}
}

@article{liuOnDemandStoragePhotonic2022,
  title = {On-{{Demand Storage}} of {{Photonic Qubits}} at {{Telecom Wavelengths}}},
  author = {Liu, Duan-Cheng and Li, Pei-Yun and Zhu, Tian-Xiang and Zheng, Liang and Huang, Jian-Yin and Zhou, Zong-Quan and Li, Chuan-Feng and Guo, Guang-Can},
  year = 2022,
  month = nov,
  journal = {Physical Review Letters},
  volume = {129},
  number = {21},
  pages = {210501},
  issn = {0031-9007, 1079-7114},
  doi = {10.1103/PhysRevLett.129.210501},
  urldate = {2024-09-08},
  langid = {english}
}

@article{moiseevEfficientMultimodeQuantum2010,
  title = {Efficient Multimode Quantum Memory Based on Photon Echo in an Optimal {{QED}} Cavity},
  author = {Moiseev, Sergey A. and Andrianov, Sergey N. and Gubaidullin, Firdus F.},
  year = 2010,
  month = aug,
  journal = {Physical Review A},
  volume = {82},
  number = {2},
  pages = {022311},
  issn = {1050-2947, 1094-1622},
  doi = {10.1103/PhysRevA.82.022311},
  urldate = {2025-05-30},
  copyright = {http://link.aps.org/licenses/aps-default-license},
  langid = {english}
}

@misc{ouMultichannelHighDimensional2025,
  title = {Multichannel and High Dimensional Integrated Photonic Quantum Memory},
  author = {Ou, Zhong-Wen and Zhu, Tian-Xiang and Liang, Peng-Jun and Hu, Xiao-Min and Zhou, Zong-Quan and Li, Chuang-Feng and Guo, Guang-Can},
  year = 2025,
  month = aug,
  eprint = {2508.19605},
  primaryclass = {quant-ph},
  publisher = {arXiv},
  urldate = {2025-08-29},
  archiveprefix = {arXiv},
  langid = {english}
}

@article{ourariIndistinguishableTelecomBand2023,
  title = {Indistinguishable Telecom Band Photons from a Single {{Er}} Ion in the Solid State},
  author = {Ourari, Salim and Dusanowski, {\L}ukasz and Horvath, Sebastian P. and Uysal, Mehmet T. and Phenicie, Christopher M. and Stevenson, Paul and Raha, Mouktik and Chen, Songtao and Cava, Robert J. and {de Leon}, Nathalie P. and Thompson, Jeff D.},
  year = 2023,
  month = aug,
  journal = {Nature},
  volume = {620},
  number = {7976},
  pages = {977--981},
  publisher = {Nature Publishing Group},
  issn = {1476-4687},
  doi = {10.1038/s41586-023-06281-4},
  urldate = {2024-10-08},
  copyright = {2023 The Author(s), under exclusive licence to Springer Nature Limited},
  langid = {english}
}

@article{rakonjacLongSpinCoherence2020,
  title = {Long Spin Coherence Times in the Ground State and in an Optically Excited State of {{Er}} 3 + 167 : {{Y}} 2 {{SiO}} 5 at Zero Magnetic Field},
  shorttitle = {Long Spin Coherence Times in the Ground State and in an Optically Excited State of {{Er}} 3 + 167},
  author = {Rakonjac, Jelena V. and Chen, Yu-Hui and Horvath, Sebastian P. and Longdell, Jevon J.},
  year = 2020,
  month = may,
  journal = {Physical Review B},
  volume = {101},
  number = {18},
  pages = {184430},
  issn = {2469-9950, 2469-9969},
  doi = {10.1103/PhysRevB.101.184430},
  urldate = {2024-11-13},
  langid = {english}
}

@article{rancicCoherenceTimeSecond2018,
  title = {Coherence Time of over a Second in a Telecom-Compatible Quantum Memory Storage Material},
  author = {Ran{\v c}i{\'c}, Milo{\v s} and Hedges, Morgan P. and Ahlefeldt, Rose L. and Sellars, Matthew J.},
  year = 2018,
  month = jan,
  journal = {Nature Physics},
  volume = {14},
  number = {1},
  pages = {50--54},
  publisher = {Nature Publishing Group},
  issn = {1745-2481},
  doi = {10.1038/nphys4254},
  urldate = {2024-08-26},
  copyright = {2017 Springer Nature Limited},
  langid = {english}
}

@article{reisererColloquiumCavityenhancedQuantum2022,
  title = {{\emph{Colloquium}} : {{Cavity-enhanced}} Quantum Network Nodes},
  shorttitle = {{\emph{Colloquium}}},
  author = {Reiserer, Andreas},
  year = 2022,
  month = dec,
  journal = {Reviews of Modern Physics},
  volume = {94},
  number = {4},
  pages = {041003},
  issn = {0034-6861, 1539-0756},
  doi = {10.1103/RevModPhys.94.041003},
  urldate = {2024-07-12},
  langid = {english}
}

@article{sabooniEfficientQuantumMemory2013,
  title = {Efficient {{Quantum Memory Using}} a {{Weakly Absorbing Sample}}},
  author = {Sabooni, Mahmood and Li, Qian and Kr{\"o}ll, Stefan and Rippe, Lars},
  year = 2013,
  month = mar,
  journal = {Physical Review Letters},
  volume = {110},
  number = {13},
  pages = {133604},
  issn = {0031-9007, 1079-7114},
  doi = {10.1103/PhysRevLett.110.133604},
  urldate = {2024-08-26},
  copyright = {http://link.aps.org/licenses/aps-default-license},
  langid = {english}
}

@article{saglamyurekBroadbandWaveguideQuantum2011,
  title = {Broadband Waveguide Quantum Memory for Entangled Photons},
  author = {Saglamyurek, Erhan and Sinclair, Neil and Jin, Jeongwan and Slater, Joshua A. and Oblak, Daniel and Bussi{\`e}res, F{\'e}lix and George, Mathew and Ricken, Raimund and Sohler, Wolfgang and Tittel, Wolfgang},
  year = 2011,
  month = jan,
  journal = {Nature},
  volume = {469},
  number = {7331},
  pages = {512--515},
  issn = {0028-0836, 1476-4687},
  doi = {10.1038/nature09719},
  urldate = {2024-11-05},
  copyright = {http://www.springer.com/tdm},
  langid = {english}
}

@article{saglamyurekMultiplexedLightmatterInterface2016,
  title = {A Multiplexed Light-Matter Interface for Fibre-Based Quantum Networks},
  author = {Saglamyurek, Erhan and Grimau Puigibert, Marcelli and Zhou, Qiang and Giner, Lambert and Marsili, Francesco and Verma, Varun B. and Woo Nam, Sae and Oesterling, Lee and Nippa, David and Oblak, Daniel and Tittel, Wolfgang},
  year = 2016,
  month = apr,
  journal = {Nature Communications},
  volume = {7},
  number = {1},
  pages = {11202},
  issn = {2041-1723},
  doi = {10.1038/ncomms11202},
  urldate = {2025-06-30},
  langid = {english}
}

@article{saglamyurekQuantumStorageEntangled2015,
  title = {Quantum Storage of Entangled Telecom-Wavelength Photons in an Erbium-Doped Optical Fibre},
  author = {Saglamyurek, Erhan and Jin, Jeongwan and Verma, Varun B. and Shaw, Matthew D. and Marsili, Francesco and Nam, Sae Woo and Oblak, Daniel and Tittel, Wolfgang},
  year = 2015,
  month = feb,
  journal = {Nature Photonics},
  volume = {9},
  number = {2},
  pages = {83--87},
  publisher = {Nature Publishing Group},
  issn = {1749-4893},
  doi = {10.1038/nphoton.2014.311},
  urldate = {2024-09-21},
  copyright = {2014 Springer Nature Limited},
  langid = {english}
}

@article{sangouardQuantumRepeatersBased2011,
  title = {Quantum Repeaters Based on Atomic Ensembles and Linear Optics},
  author = {Sangouard, Nicolas and Simon, Christoph and De Riedmatten, Hugues and Gisin, Nicolas},
  year = 2011,
  month = mar,
  journal = {Reviews of Modern Physics},
  volume = {83},
  number = {1},
  pages = {33--80},
  issn = {0034-6861, 1539-0756},
  doi = {10.1103/RevModPhys.83.33},
  urldate = {2024-10-22},
  copyright = {http://link.aps.org/licenses/aps-default-license},
  langid = {english}
}

@article{seriQuantumStorageFrequencyMultiplexed2019,
  title = {Quantum {{Storage}} of {{Frequency-Multiplexed Heralded Single Photons}}},
  author = {Seri, Alessandro and {Lago-Rivera}, Dario and Lenhard, Andreas and Corrielli, Giacomo and Osellame, Roberto and Mazzera, Margherita and {de Riedmatten}, Hugues},
  year = 2019,
  month = aug,
  journal = {Physical Review Letters},
  volume = {123},
  number = {8},
  pages = {080502},
  publisher = {American Physical Society},
  doi = {10.1103/PhysRevLett.123.080502},
  urldate = {2024-06-03},
  langid = {american}
}

@article{sinclairSpectralMultiplexingScalable2014,
  title = {Spectral {{Multiplexing}} for {{Scalable Quantum Photonics}} Using an {{Atomic Frequency Comb Quantum Memory}} and {{Feed-Forward Control}}},
  author = {Sinclair, Neil and Saglamyurek, Erhan and Mallahzadeh, Hassan and Slater, Joshua A. and George, Mathew and Ricken, Raimund and Hedges, Morgan P. and Oblak, Daniel and Simon, Christoph and Sohler, Wolfgang and Tittel, Wolfgang},
  year = 2014,
  month = jul,
  journal = {Physical Review Letters},
  volume = {113},
  number = {5},
  pages = {053603},
  issn = {0031-9007, 1079-7114},
  doi = {10.1103/PhysRevLett.113.053603},
  urldate = {2024-08-22},
  copyright = {http://link.aps.org/licenses/aps-default-license},
  langid = {english}
}

@article{stuartInitializationProtocolEfficient2021,
  title = {Initialization Protocol for Efficient Quantum Memories Using Resolved Hyperfine Structure},
  author = {Stuart, James S. and Hedges, Morgan and Ahlefeldt, Rose and Sellars, Matthew},
  year = 2021,
  month = aug,
  journal = {Physical Review Research},
  volume = {3},
  number = {3},
  pages = {L032054},
  issn = {2643-1564},
  doi = {10.1103/PhysRevResearch.3.L032054},
  urldate = {2025-06-14},
  langid = {english}
}

@article{tellerQuantumStorageQubits2025,
  title = {Quantum {{Storage}} of {{Qubits}} in an {{Array}} of {{Independently Controllable Solid-State Quantum Memories}}},
  author = {Teller, Markus and Plascencia, Susana and Grandi, Samuele and de Riedmatten, Hugues},
  year = 2025,
  month = aug,
  journal = {Physical Review X},
  volume = {15},
  number = {3},
  eprint = {2509.11910},
  primaryclass = {quant-ph},
  pages = {031053},
  issn = {2160-3308},
  doi = {10.1103/z6lc-qw2d},
  urldate = {2025-09-17},
  archiveprefix = {arXiv},
  langid = {english}
}

@article{thielOpticalDecoherencePersistent2010,
  title = {Optical Decoherence and Persistent Spectral Hole Burning in {{Er3}}+:{{LiNbO3}}},
  shorttitle = {Optical Decoherence and Persistent Spectral Hole Burning in {{Er3}}+},
  author = {Thiel, C.W. and Macfarlane, R.M. and B{\"o}ttger, T. and Sun, Y. and Cone, R.L. and Babbitt, W.R.},
  year = 2010,
  month = sep,
  journal = {Journal of Luminescence},
  volume = {130},
  number = {9},
  pages = {1603--1609},
  issn = {00222313},
  doi = {10.1016/j.jlumin.2009.12.020},
  urldate = {2024-11-11},
  copyright = {https://www.elsevier.com/tdm/userlicense/1.0/},
  langid = {english}
}

@article{thielRareearthdopedMaterialsApplications2011,
  title = {Rare-Earth-Doped Materials for Applications in Quantum Information Storage and Signal Processing},
  author = {Thiel, C.W. and B{\"o}ttger, Thomas and Cone, R.L.},
  year = 2011,
  month = mar,
  journal = {Journal of Luminescence},
  volume = {131},
  number = {3},
  pages = {353--361},
  issn = {00222313},
  doi = {10.1016/j.jlumin.2010.12.015},
  urldate = {2024-11-11},
  copyright = {https://www.elsevier.com/tdm/userlicense/1.0/},
  langid = {english}
}

@article{ulanowskiSpectralMultiplexingTelecom2022,
  title = {Spectral Multiplexing of Telecom Emitters with Stable Transition Frequency},
  author = {Ulanowski, Alexander and Merkel, Benjamin and Reiserer, Andreas},
  year = 2022,
  month = oct,
  journal = {Science Advances},
  volume = {8},
  number = {43},
  pages = {eabo4538},
  issn = {2375-2548},
  doi = {10.1126/sciadv.abo4538},
  urldate = {2024-07-12},
  langid = {english}
}

@article{yangControllingSingleRare2023,
  title = {Controlling Single Rare Earth Ion Emission in an Electro-Optical Nanocavity},
  author = {Yang, Likai and Wang, Sihao and Shen, Mohan and Xie, Jiacheng and {Hong X. Tang}},
  year = 2023,
  month = mar,
  journal = {Nature Communications},
  volume = {14},
  number = {1},
  pages = {1718},
  publisher = {Nature Publishing Group},
  issn = {2041-1723},
  doi = {10.1038/s41467-023-37513-w},
  urldate = {2024-09-18},
  copyright = {2023 The Author(s)},
  langid = {english}
}

@article{yuFrequencyTunableCavityEnhanced2023,
  title = {Frequency {{Tunable}}, {{Cavity-Enhanced Single Erbium Quantum Emitter}} in the {{Telecom Band}}},
  author = {Yu, Yong and Oser, Dorian and Da Prato, Gaia and Urbinati, Emanuele and {\'A}vila, Javier Carrasco and Zhang, Yu and Remy, Patrick and Marzban, Sara and Gr{\"o}blacher, Simon and Tittel, Wolfgang},
  year = 2023,
  month = oct,
  journal = {Physical Review Letters},
  volume = {131},
  number = {17},
  pages = {170801},
  issn = {0031-9007, 1079-7114},
  doi = {10.1103/PhysRevLett.131.170801},
  urldate = {2024-07-12},
  langid = {english}
}

@article{zhongOpticallyAddressableNuclear2015,
  title = {Optically Addressable Nuclear Spins in a Solid with a Six-Hour Coherence Time},
  author = {Zhong, Manjin and Hedges, Morgan P. and Ahlefeldt, Rose L. and Bartholomew, John G. and Beavan, Sarah E. and Wittig, Sven M. and Longdell, Jevon J. and Sellars, Matthew J.},
  year = 2015,
  month = jan,
  journal = {Nature},
  volume = {517},
  number = {7533},
  pages = {177--180},
  publisher = {Nature Publishing Group},
  issn = {1476-4687},
  doi = {10.1038/nature14025},
  urldate = {2024-08-28},
  copyright = {2015 Springer Nature Limited},
  langid = {english}
}

@article{zhouPhotonicIntegratedQuantum2023,
  title = {Photonic {{Integrated Quantum Memory}} in {{Rare-Earth Doped Solids}}},
  author = {Zhou, Zong-Quan and Liu, Chao and Li, Chuan-Feng and Guo, Guang-Can and Oblak, Daniel and Lei, Mi and Faraon, Andrei and Mazzera, Margherita and {de Riedmatten}, Hugues},
  year = 2023,
  journal = {Laser \& Photonics Reviews},
  volume = {17},
  number = {10},
  pages = {2300257},
  issn = {1863-8899},
  doi = {10.1002/lpor.202300257},
  urldate = {2024-10-08},
  langid = {english}
}

@article{boesLithiumNiobatePhotonics2023,
  title = {Lithium Niobate Photonics: {{Unlocking}} the Electromagnetic Spectrum},
  shorttitle = {Lithium Niobate Photonics},
  author = {Boes, Andreas and Chang, Lin and Langrock, Carsten and Yu, Mengjie and Zhang, Mian and Lin, Qiang and Lon{\v c}ar, Marko and Fejer, Martin and Bowers, John and Mitchell, Arnan},
  year = 2023,
  month = jan,
  journal = {Science},
  volume = {379},
  number = {6627},
  pages = {eabj4396},
  issn = {0036-8075, 1095-9203},
  doi = {10.1126/science.abj4396},
  urldate = {2024-03-30},
  langid = {english}
}

@article{colangeloMolybdenumSilicideSuperconducting2024,
  title = {Molybdenum {{Silicide Superconducting Nanowire Single-Photon Detectors}} on {{Lithium Niobate Waveguides}}},
  author = {Colangelo, Marco and Zhu, Di and Shao, Linbo and Holzgrafe, Jeffrey and Batson, Emma K. and Desiatov, Boris and Medeiros, Owen and Yeung, Matthew and Loncar, Marko and Berggren, Karl K.},
  year = 2024,
  month = feb,
  journal = {ACS Photonics},
  volume = {11},
  number = {2},
  pages = {356--361},
  issn = {2330-4022, 2330-4022},
  doi = {10.1021/acsphotonics.3c01628},
  urldate = {2024-03-30},
  copyright = {7.300},
  langid = {english},
  lccn = {7.000}
}

@article{lomonteSinglephotonDetectionCryogenic2021,
  title = {Single-Photon Detection and Cryogenic Reconfigurability in Lithium Niobate Nanophotonic Circuits},
  author = {Lomonte, Emma and Wolff, Martin A. and Beutel, Fabian and Ferrari, Simone and Schuck, Carsten and Pernice, Wolfram H. P. and Lenzini, Francesco},
  year = 2021,
  month = nov,
  journal = {Nature Communications},
  volume = {12},
  number = {1},
  pages = {6847},
  issn = {2041-1723},
  doi = {10.1038/s41467-021-27205-8},
  urldate = {2024-03-30},
  copyright = {17.000},
  langid = {english},
  lccn = {16.600}
}

@article{pelucchiPotentialGlobalOutlook2021,
  title = {The Potential and Global Outlook of Integrated Photonics for Quantum Technologies},
  author = {Pelucchi, Emanuele and Fagas, Giorgos and Aharonovich, Igor and Englund, Dirk and Figueroa, Eden and Gong, Qihuang and Hannes, H{\"u}bel and Liu, Jin and Lu, Chao-Yang and Matsuda, Nobuyuki and Pan, Jian-Wei and Schreck, Florian and Sciarrino, Fabio and Silberhorn, Christine and Wang, Jianwei and J{\"o}ns, Klaus D.},
  year = 2021,
  month = dec,
  journal = {Nature Reviews Physics},
  volume = {4},
  number = {3},
  pages = {194--208},
  issn = {2522-5820},
  doi = {10.1038/s42254-021-00398-z},
  urldate = {2024-03-30},
  langid = {english}
}

@article{wangIntegratedLithiumNiobate2018,
  title = {Integrated Lithium Niobate Electro-Optic Modulators Operating at {{CMOS-compatible}} Voltages},
  author = {Wang, Cheng and Zhang, Mian and Chen, Xi and Bertrand, Maxime and {Shams-Ansari}, Amirhassan and Chandrasekhar, Sethumadhavan and Winzer, Peter and Lon{\v c}ar, Marko},
  year = 2018,
  month = oct,
  journal = {Nature},
  volume = {562},
  number = {7725},
  pages = {101--104},
  issn = {0028-0836, 1476-4687},
  doi = {10.1038/s41586-018-0551-y},
  urldate = {2024-03-30},
  langid = {english}
}

@article{warnerDCstableThinfilmLithium2025,
  title = {{{DC-stable}} Thin-Film Lithium Niobate Modulator at Liquid Nitrogen Temperatures},
  author = {Warner, Hana K. and Zhao, Yuqi and Zhang, Yu and Zhang, Mian and Lon{\v c}ar, Marko},
  year = 2025,
  month = sep,
  journal = {Optics Letters},
  volume = {50},
  number = {17},
  pages = {5398},
  issn = {0146-9592, 1539-4794},
  doi = {10.1364/OL.571632},
  urldate = {2025-09-01},
  langid = {english}
}

@article{xiaTunableMicrocavitiesCoupled2022,
  title = {Tunable Microcavities Coupled to Rare-Earth Quantum Emitters},
  author = {Xia, Kangwei and Sardi, Fiammetta and Sauerzapf, Colin and Kornher, Thomas and Becker, Hans-Werner and Kis, Zsolt and Kovacs, Laszlo and Dertli, Denis and Foglszinger, Jonas and Kolesov, Roman and Wrachtrup, J{\"o}rg},
  year = 2022,
  month = apr,
  journal = {Optica},
  volume = {9},
  number = {4},
  pages = {445--450},
  publisher = {Optica Publishing Group},
  issn = {2334-2536},
  doi = {10.1364/OPTICA.453527},
  urldate = {2024-04-08},
  copyright = {\copyright{} 2022 Optica Publishing Group},
  langid = {english}
}

@article{kimbleQuantumInternet2008,
  title = {The Quantum Internet},
  author = {Kimble, H. J.},
  year = 2008,
  month = jun,
  journal = {Nature},
  volume = {453},
  number = {7198},
  pages = {1023--1030},
  issn = {0028-0836, 1476-4687},
  doi = {10.1038/nature07127},
  urldate = {2026-04-20},
  copyright = {http://www.springer.com/tdm},
  langid = {english}
}

@article{lauritzenTelecommunicationWavelengthSolidStateMemory2010,
  title = {Telecommunication-{{Wavelength Solid-State Memory}} at the {{Single Photon Level}}},
  author = {Lauritzen, Bj{\"o}rn and Min{\'a}{\v r}, Ji{\v r}{\'i} and De Riedmatten, Hugues and Afzelius, Mikael and Sangouard, Nicolas and Simon, Christoph and Gisin, Nicolas},
  year = 2010,
  month = feb,
  journal = {Physical Review Letters},
  volume = {104},
  number = {8},
  pages = {080502},
  issn = {0031-9007, 1079-7114},
  doi = {10.1103/PhysRevLett.104.080502},
  urldate = {2026-04-20},
  copyright = {http://link.aps.org/licenses/aps-default-license},
  langid = {english}
}

@article{anQuantumTeleportationTelecom2025,
  title = {Quantum {{Teleportation}} from {{Telecom Photons}} to {{Erbium-Ion Ensembles}}},
  author = {An, Yu-Yang and He, Qian and Xue, Wenyi and Jiang, Ming-Hao and Yang, Chengdong and Lu, Yan-Qing and Zhu, Shining and Ma, Xiao-Song},
  year = 2025,
  month = jul,
  journal = {Physical Review Letters},
  volume = {135},
  number = {1},
  pages = {010804},
  issn = {0031-9007, 1079-7114},
  doi = {10.1103/3wh8-2gh1},
  urldate = {2026-04-20},
  langid = {english}
}

@article{
zhangTelecombandintegratedMultimodePhotonic2023,
author = {Xueying Zhang  and Bin Zhang  and Shihai Wei  and Hao Li  and Jinyu Liao  and Cheng Li  and Guangwei Deng  and You Wang  and Haizhi Song  and Lixing You  and Bo Jing  and Feng Chen  and Guangcan Guo  and Qiang Zhou },
title = {Telecom-band–integrated multimode photonic quantum memory},
journal = {Science Advances},
volume = {9},
number = {28},
pages = {eadf4587},
year = {2023},
doi = {10.1126/sciadv.adf4587}}

@article{durantiEfficientCavityassistedStorage2024,
  title = {Efficient Cavity-Assisted Storage of Photonic Qubits in a Solid-State Quantum Memory},
  author = {Duranti, Stefano and Wengerowsky, S{\"o}ren and Feldmann, Leo and Seri, Alessandro and Casabone, Bernardo and De Riedmatten, Hugues},
  year = 2024,
  month = jul,
  journal = {Optics Express},
  volume = {32},
  number = {15},
  pages = {26884},
  issn = {1094-4087},
  doi = {10.1364/OE.512318},
  urldate = {2026-04-20},
  langid = {english}
}

@article{sinclairPropertiesRareEarthIonDopedWaveguide2017,
  title = {Properties of a {{Rare-Earth-Ion-Doped Waveguide}} at {{Sub-Kelvin Temperatures}} for {{Quantum Signal Processing}}},
  author = {Sinclair, N. and Oblak, D. and Thiel, C. W. and Cone, R. L. and Tittel, W.},
  year = 2017,
  month = mar,
  journal = {Physical Review Letters},
  volume = {118},
  number = {10},
  pages = {100504},
  issn = {0031-9007, 1079-7114},
  doi = {10.1103/PhysRevLett.118.100504},
  urldate = {2026-04-20},
  copyright = {http://link.aps.org/licenses/aps-default-license},
  langid = {english}
}

@article{zhuTwentynineMillionIntrinsic2024b,
  title = {Twenty-Nine Million Intrinsic {{{\emph{Q}}}} -Factor Monolithic Microresonators on Thin-Film Lithium Niobate},
  author = {Zhu, Xinrui and Hu, Yaowen and Lu, Shengyuan and Warner, Hana K. and Li, Xudong and Song, Yunxiang and Magalh{\~a}es, Let{\'i}cia and {Shams-Ansari}, Amirhassan and Cordaro, Andrea and Sinclair, Neil and Lon{\v c}ar, Marko},
  year = 2024,
  month = aug,
  journal = {Photonics Research},
  volume = {12},
  number = {8},
  pages = {A63},
  issn = {2327-9125},
  doi = {10.1364/PRJ.521172},
  urldate = {2026-04-20},
  langid = {english}
}

@article{baryaUltraHighQTunable2025a,
  title = {Ultra High-{{Q}} Tunable Microring Resonators Enabled by Slow Light},
  author = {Barya, Priyash and Prabhu, Ashwith and Heller, Laura and Chow, Edmond and Goldschmidt, Elizabeth A.},
  year = 2025,
  month = nov,
  journal = {Nature Communications},
  volume = {16},
  number = {1},
  pages = {10496},
  issn = {2041-1723},
  doi = {10.1038/s41467-025-65533-1},
  urldate = {2026-04-20},
  langid = {english}
}

@article{puExperimentalRealizationMultiplexed2017,
  title = {Experimental Realization of a Multiplexed Quantum Memory with 225 Individually Accessible Memory Cells},
  author = {Pu, Y-F and Jiang, N. and Chang, W. and Yang, H-X and Li, C. and Duan, L-M},
  year = 2017,
  month = may,
  journal = {Nature Communications},
  volume = {8},
  number = {1},
  pages = {15359},
  issn = {2041-1723},
  doi = {10.1038/ncomms15359},
  urldate = {2026-04-21},
  langid = {english}
}

@article{langeWidelyNondegenerateNonlinear2025,
  title = {Widely Non-Degenerate Nonlinear Frequency Conversion in Cryogenic Titanium in-Diffused Lithium Niobate Waveguides},
  author = {Lange, Nina Amelie and Lengeling, Sebastian and Mues, Philipp and Quiring, Viktor and Ridder, Werner and Eigner, Christof and Herrmann, Harald and Silberhorn, Christine and Bartley, Tim J.},
  year = 2025,
  month = dec,
  journal = {Optics Express},
  volume = {33},
  number = {24},
  pages = {50451},
  issn = {1094-4087},
  doi = {10.1364/OE.578108},
  urldate = {2026-04-21},
  langid = {english}
}

@article{lutzEffectsMechanicalProcessing2017,
  title = {Effects of Mechanical Processing and Annealing on Optical Coherence Properties of {{Er}} 3 + :{{LiNbO3}} Powders},
  shorttitle = {Effects of Mechanical Processing and Annealing on Optical Coherence Properties of {{Er}} 3 +},
  author = {Lutz, Thomas and Veissier, Lucile and Thiel, Charles W. and Woodburn, Philip J.T. and Cone, Rufus L. and Barclay, Paul E. and Tittel, Wolfgang},
  year = 2017,
  month = nov,
  journal = {Journal of Luminescence},
  volume = {191},
  pages = {2--12},
  issn = {00222313},
  doi = {10.1016/j.jlumin.2017.03.027},
  urldate = {2026-04-21},
  langid = {english}
}

@article{zhaoHighQualityEntangled2020,
  title = {High {{Quality Entangled Photon Pair Generation}} in {{Periodically Poled Thin-Film Lithium Niobate Waveguides}}},
  author = {Zhao, Jie and Ma, Chaoxuan and R{\"u}sing, Michael and Mookherjea, Shayan},
  year = 2020,
  month = apr,
  journal = {Physical Review Letters},
  volume = {124},
  number = {16},
  pages = {163603},
  issn = {0031-9007, 1079-7114},
  doi = {10.1103/PhysRevLett.124.163603},
  urldate = {2024-03-30},
  langid = {english}
}

@article{fransonBellInequalityPosition1989,
  title = {Bell Inequality for Position and Time},
  author = {Franson, J. D.},
  year = 1989,
  month = may,
  journal = {Physical Review Letters},
  volume = {62},
  number = {19},
  pages = {2205--2208},
  issn = {0031-9007},
  doi = {10.1103/PhysRevLett.62.2205},
  urldate = {2026-04-22},
  copyright = {http://link.aps.org/licenses/aps-default-license},
  langid = {english}
}

@article{wehnerQuantumInternetVision2018,
  title = {Quantum Internet: {{A}} Vision for the Road Ahead},
  shorttitle = {Quantum Internet},
  author = {Wehner, Stephanie and Elkouss, David and Hanson, Ronald},
  year = 2018,
  month = oct,
  journal = {Science},
  volume = {362},
  number = {6412},
  pages = {eaam9288},
  issn = {0036-8075, 1095-9203},
  doi = {10.1126/science.aam9288},
  urldate = {2026-04-22},
}

@article{briegelQuantumRepeatersRole1998,
  title = {Quantum {{Repeaters}}: {{The Role}} of {{Imperfect Local Operations}} in {{Quantum Communication}}},
  shorttitle = {Quantum {{Repeaters}}},
  author = {Briegel, H.-J. and D{\"u}r, W. and Cirac, J. I. and Zoller, P.},
  year = 1998,
  month = dec,
  journal = {Physical Review Letters},
  volume = {81},
  number = {26},
  pages = {5932--5935},
  issn = {0031-9007, 1079-7114},
  doi = {10.1103/PhysRevLett.81.5932},
  urldate = {2026-04-22},
  copyright = {http://link.aps.org/licenses/aps-default-license},
  langid = {english}
}

@article{lvovskyOpticalQuantumMemory2009,
  title = {Optical Quantum Memory},
  author = {Lvovsky, Alexander I. and Sanders, Barry C. and Tittel, Wolfgang},
  year = 2009,
  month = dec,
  journal = {Nature Photonics},
  volume = {3},
  number = {12},
  pages = {706--714},
  issn = {1749-4885, 1749-4893},
  doi = {10.1038/nphoton.2009.231},
  urldate = {2026-04-22},
  copyright = {http://www.springer.com/tdm},
  langid = {english}
}

@article{maOnehourCoherentOptical2021,
  title = {One-Hour Coherent Optical Storage in an Atomic Frequency Comb Memory},
  author = {Ma, Yu and Ma, You-Zhi and Zhou, Zong-Quan and Li, Chuan-Feng and Guo, Guang-Can},
  year = 2021,
  month = apr,
  journal = {Nature Communications},
  volume = {12},
  number = {1},
  pages = {2381},
  issn = {2041-1723},
  doi = {10.1038/s41467-021-22706-y},
  urldate = {2026-04-22},
  langid = {english}
}

@article{leiQuantumOpticalMemory2023,
  title = {Quantum Optical Memory for Entanglement Distribution},
  author = {Lei, Yisheng and Kimiaee Asadi, Faezeh and Zhong, Tian and Kuzmich, Alex and Simon, Christoph and Hosseini, Mahdi},
  year = 2023,
  month = nov,
  journal = {Optica},
  volume = {10},
  number = {11},
  pages = {1511},
  issn = {2334-2536},
  doi = {10.1364/OPTICA.493732},
  urldate = {2026-04-26},
  langid = {english}
}

@article{liEfficientStorageMultidimensional2025,
  title = {Efficient Storage of Multidimensional Telecom Photons in a Solid-State Quantum Memory},
  author = {Li, Zongfeng and Lei, Yisheng and Kling, Trevor and Hosseini, Mahdi},
  year = 2025,
  month = jun,
  journal = {Optica Quantum},
  volume = {3},
  number = {3},
  pages = {295},
  issn = {2837-6714},
  doi = {10.1364/OPTICAQ.564321},
  urldate = {2026-04-26},
  langid = {english}
}

@misc{ulanowskiCavityenhancedOpticalReadout2026,
  title = {Cavity-Enhanced Optical Readout and Control of Nuclear Spin Qubits},
  author = {Ulanowski, Alexander and Fr{\"u}h, Johannes and Salamon, Fabian and Holz{\"a}pfel, Adrian and Reiserer, Andreas},
  year = 2026,
  month = mar,
  eprint = {2603.01987},
  primaryclass = {quant-ph},
  publisher = {arXiv},
  urldate = {2026-04-26},
  archiveprefix = {arXiv},
  langid = {english}
}

@article{mengEfficientIntegratedQuantum2026,
  title = {Efficient Integrated Quantum Memory for Light},
  author = {Meng, Ruo-Ran and Liu, Pei-Xi and Liu, Xiao and Zhu, Tian-Xiang and Liang, Peng-Jun and Zhang, Chao and Tang, Zhong-Yang and Zhang, Hong-Zhe and Cui, Jin-Ming and Jin, Ming and Zhou, Zong-Quan and Li, Chuan-Feng and Guo, Guang-Can},
  year = 2026,
  month = apr,
  journal = {Nature Photonics},
  volume = {20},
  number = {4},
  pages = {437--443},
  issn = {1749-4885, 1749-4893},
  doi = {10.1038/s41566-026-01845-y},
  urldate = {2026-04-26},
  langid = {english}
}

@article{huangStarkTuningTelecom2023,
  title = {Stark {{Tuning}} of {{Telecom Single-Photon Emitters Based}} on a {{Single Er}}{\textsuperscript{3+}}},
  author = {Huang, Jian-Yin and Liang, Peng-Jun and Zheng, Liang and Li, Pei-Yun and Ma, You-Zhi and Liu, Duan-Chen and Xie, Jing-Hui and Zhou, Zong-Quan and Li, Chuan-Feng and Guo, Guang-Can},
  year = 2023,
  month = jun,
  journal = {Chinese Physics Letters},
  volume = {40},
  number = {7},
  pages = {070301},
  issn = {0256-307X, 1741-3540},
  doi = {10.1088/0256-307X/40/7/070301},
  urldate = {2026-04-27},
  langid = {english}
}

@article{dikandeContinuouswavePulseRegimes2017,
  title = {Continuous-Wave to Pulse Regimes for a Family of Passively Mode-Locked Lasers with Saturable Nonlinearity},
  author = {Dikand{\'e}, Alain M and Titafan, J Voma and Essimbi, B Z},
  year = 2017,
  month = oct,
  journal = {Journal of Optics},
  volume = {19},
  number = {10},
  pages = {105504},
  issn = {2040-8978, 2040-8986},
  doi = {10.1088/2040-8986/aa8255},
  urldate = {2026-04-28},
  langid = {english}
}

@book{bevington2003data,
  title={Data Reduction and Error Analysis for the Physical Sciences},
  author={Bevington, P. and Robinson, D.K.},
  isbn={9780072472271},
  lccn={91027275},
  year={2003},
  publisher={McGraw-Hill Education}
}

@article{duttaAtomicFrequencyComb2023a,
  title = {An {{Atomic Frequency Comb Memory}} in {{Rare-Earth-Doped Thin-Film Lithium Niobate}}},
  author = {Dutta, Subhojit and Zhao, Yuqi and Saha, Uday and Farfurnik, Demitry and Goldschmidt, Elizabeth A. and Waks, Edo},
  year = 2023,
  month = apr,
  journal = {ACS Photonics},
  volume = {10},
  number = {4},
  pages = {1104--1109},
  publisher = {American Chemical Society},
  doi = {10.1021/acsphotonics.2c01835}
}

@article{sayemLithiumniobateoninsulatorWaveguideintegratedSuperconducting2020,
  title = {Lithium-Niobate-on-Insulator Waveguide-Integrated Superconducting Nanowire Single-Photon Detectors},
  author = {Sayem, Ayed Al and Cheng, Risheng and Wang, Sihao and Tang, Hong X.},
  year = 2020,
  month = apr,
  journal = {Applied Physics Letters},
  volume = {116},
  number = {15},
  pages = {151102},
  issn = {0003-6951},
  doi = {10.1063/1.5142852},
  urldate = {2026-09-05}
}

@article{gritschOpticalSingleshotReadout2025,
  title = {Optical Single-Shot Readout of Spin Qubits in Silicon},
  author = {Gritsch, Andreas and Ulanowski, Alexander and Pforr, Jakob and Reiserer, Andreas},
  year = 2025,
  month = jan,
  journal = {Nature Communications},
  volume = {16},
  number = {1},
  pages = {64},
  issn = {2041-1723},
  doi = {10.1038/s41467-024-55552-9}
}

@article{uysalSpinPhotonEntanglementSingle2025a,
  title = {Spin-Photon Entanglement of a Single ${\mathrm{Er}}^{3+}$ Ion in the Telecom Band},
  author = {Uysal, Mehmet T. and Dusanowski, {\L}ukasz and Xu, Haitong and Horvath, Sebastian P. and Ourari, Salim and Cava, Robert J. and {de Leon}, Nathalie P. and Thompson, Jeff D.},
  year = 2025,
  month = mar,
  journal = {Physical Review X},
  volume = {15},
  number = {1},
  pages = {011071},
  publisher = {American Physical Society},
  doi = {10.1103/PhysRevX.15.011071}
}

@article{zhongNanophotonicRareearthQuantum2017a,
  title = {Nanophotonic Rare-Earth Quantum Memory with Optically Controlled Retrieval},
  author = {Zhong, Tian and Kindem, Jonathan M. and Bartholomew, John G. and Rochman, Jake and Craiciu, Ioana and Miyazono, Evan and Bettinelli, Marco and Cavalli, Enrico and Verma, Varun and Nam, Sae Woo and Marsili, Francesco and Shaw, Matthew D. and Beyer, Andrew D. and Faraon, Andrei},
  year = 2017,
  month = sep,
  journal = {Science},
  volume = {357},
  number = {6358},
  pages = {1392--1395},
  publisher = {American Association for the Advancement of Science},
  doi = {10.1126/science.aan5959},
  urldate = {2026-05-08}
}

\end{document}